\documentclass[useAMS,usenatbib]{mn2e}

\usepackage{graphicx}
\usepackage{ifthen}
\usepackage{multirow} 

\newcommand{\gal}[1]{\ifthenelse{\equal{#1}{1}}{\citet{gallazzi-2005-362}}{G05}}
\newcommand{\kau}[1]{\ifthenelse{\equal{#1}{1}}{\citet{kauffmann-2003-341}}{K03}}

\newcommand{\Msun}{\mathop{\rm M_{\odot}\,}\nolimits}

\title[Properties of LRGs in the SDSS]{Properties of Luminous Red Galaxies in the Sloan Digital Sky Survey}
\author[Tom Barber, Avery Meiksin and Tara Murphy]
{Tom Barber$^{1}$\thanks{E-mail: tjb@roe.ac.uk (TJB);
aam@roe.ac.uk (AM); tara@physics.usyd.edu.au (TM)},
Avery Meiksin$^{1}$\footnotemark[1] and Tara Murphy$^{2}$\footnotemark[1]\\
{$^{1}$ SUPA\thanks{Scottish Universities Physics Alliance};
Institute for Astronomy, University of Edinburgh, Royal Observatory,
Edinburgh, EH9 3HJ}\\
{$^{2}$ School of Physics and School of Information Technologies, University of Sydney, NSW 2006, Australia}}

\begin{document}


\pagerange{\pageref{firstpage}--\pageref{lastpage}} \pubyear{2006}

\maketitle

\label{firstpage}

\begin{abstract}
We perform population synthesis modelling of a magnitude-limited
sample of 4391 Luminous Red Galaxies selected from the Sloan Digital
Sky Survey Data Release 4 (SDSS DR4). We fit measured spectral indices
using a large library of high resolution spectra, covering a wide
range of metallicities and assuming an exponentially decaying
star-formation rate punctuated by bursts, to obtain median-likelihood
estimates for the light-weighted age, metallicity, stellar mass and
extinction for the galaxies. The ages lie predominantly in the range
4--10~Gyr, peaking near 6~Gyr, with metallicities in the range
$-0.4<{\rm [Z/H]}<0.4$, peaking at ${\rm [Z/H]}\approx0.2$. Only a few
per cent of the spectra are better fit allowing for a burst in
addition to continuous star-formation. For these systems, typically
one-quarter to one-third of the stars are formed in the burst. The
total stellar masses of all the galaxies are confined to a very narrow
range around $\sim3\times10^{11}\,M_\odot$, consistent with the
expected homogeneity of the sample. Our results broadly agree with
those of previous groups using an independent population synthesis
code. We find, however, that our choice in priors results in ages
1--2~Gyr smaller, decreasing the peak formation epoch from about
$z=2.3$ to $z=1.3$ for the stars. To describe the distribution in
measured mean metallicity of the galaxies, we develop a metal
evolution model incorporating stochastic star-formation quenching
motivated by recent attempts to account for the apparent
`anti-hierarchical' formation of elliptical galaxies. Two scenarios
emerge, a closed box with an effective stellar yield of 0.26, and an
accreting box with an effective stellar yield of 0.10. Both scenarios
require an IMF weighted towards massive stars. They also require
characteristic star-formation quenching times of about $10^8$~yr, the
expected lifetime of luminous Quasi-Stellar Objects. The models
predict an anti-correlation between the age and mean metallicity of
the galaxies similar to that observed.

\end{abstract}


\section{Introduction}

In the hierarchical merger scenario of structure formation, more
massive systems form from the merger of less massive ones
\citep{1968ApJ...154..891P, 1983ApJ...274....1P}. While numerical
simulations of the evolution of dark matter haloes in a Cold Dark
Matter dominated universe bear this out \citep{1985ApJ...292..371D},
predictions in the hierarchical scenario for galaxies are complicated
by the need to account for gaseous dissipation
\citep{1978MNRAS.183..341W}, without which galaxies would lose their
integrity at each level of the merger hierarchy. The most
straightforward prediction of the hierarchical scenario is that
smaller mass galaxies form prior to the more massive, although the
effects of galaxy mergers, as distinct from halo mergers, could
substantially complicate this picture
\citep{1984Natur.311..517B}. Early semi-analytic estimates of the
formation and properties of galaxies have suggested that half of the
stars in the Universe formed at redshifts $z<1.5$, and only some 10
per cent at $z>3$ \citep{1994MNRAS.271..781C, 1998ApJ...498..504B}.
Elliptical galaxies in particular are predicted to show a trend of
decreasing age with increasing luminosity, with the brightest
ellipticals the youngest \citep{1996MNRAS.281..487K, 1998MNRAS.294..705K}.

Observations have generally not supported the predicted ages and their
trends with galaxy properties. The contrary evidence is derived
primarily from two methods: measurements of the luminosity function of
galaxies and its evolution, and age estimates based on population
synthesis. \citet{1996AJ....112..839C} discovered that the
characteristic luminosity of galaxies undergoing rapid star formation
has been declining with time at redshifts $z<1.7$, a result extended
to $z\approx3$ by \citet{1997AJ....113....1S}. As a function of galaxy
class, the later the type of galaxy, the later is its peak period of
star formation \citep{1997MNRAS.285..613H}. Since then, an abundance
of evidence for anti-hierarchical galaxy formation (or so-called
`downsizing') has accrued based on galaxy number counts and colours,
generally indicating that the `quenching mass', at which the major
episode of star formation ceases, decreases with decreasing redshift
\citep{2003A&A...402..837P, 2004A&A...424...23F, 2005ApJ...625..621B,
2005ApJ...619L.131D, Bundy05, Cimatti06}. Massive galaxies appear to
undergo a higher merger rate at high redshifts ($z\approx3$) than at
low redshifts, settling down by $z\approx1.5$
\citep{2006ApJ...638..686C}, again in apparent opposition to the
hierarchical merger scenario.

The picture that appears to be developing is that star formation
occurred early in systems with a wide range of masses, but was
terminated through some unknown process, first in the progenitors of
the most massive galaxies, and progressing steadily towards less
massive systems with time. The progenitors of massive galaxies (with
stellar masses today in excess of $10^{10}\,\Msun$), subsequently
merged, possibly with a small amount of additional star formation, but
settled into largely quiescent systems by $z=1.5$.

Numerical modellers have sought to catch up with these findings,
drawing increasingly on the role of feedback. While supernova feedback
has long been assigned a key role in regulating star formation, Active
Galactic Nuclei have now been invoked as a means of terminating star
formation in the most massive systems, and proposed as the origin of
the dichotomy between blue galaxies and the more massive red
\citep{2006MNRAS.368....2D}. Incorporating feedback by supernovae and
Active Galactic Nuclei has become a promissing means of reconciling
the mass-sensitive evolution in number counts, the merger rates, and
the ages of stellar populations \citep{2006MNRAS.370..645B,
DeLucia06}. These models may be expected to develop further as
knowledge of the star formation history of galaxies of different
masses and in different environments continues to grow increasingly
precise.

In the past, population synthesis model estimates of ages were long
plagued by the age-metallicity degeneracy. Improvements in the stellar
atmosphere models, the data, and the understanding of the sensitivity
of absorption lines to age and metallicity have since led to more
reliable age and metallicity determinations. \citet{trager-2000-119,
2000AJ....120..165T} found a range of ages for elliptical galaxies in
the field, with lower velocity-dispersion galaxies tending to be
younger. By contrast, ellipticals in clusters tend to be old. They
also found that larger velocity-dispersion galaxies tend to show
increased suppression of iron-peak elements. \citet{caldwell-2003-125}
similarly found that smaller velocity dispersion systems tend to have
younger luminosity-weighted ages. Evidence for recent burst activity
in field ellipticals was found by \citet{2002ApJ...564L..13T}.
\citet{thomas-2005-621} have suggested a shift in the peak star
formation epochs of early type galaxies, with massive early-type
galaxies in low-density environments $\sim2$~Gyr younger than their
high-density environment counterparts, and the star formation rate
peak shifting to lower redshifts the lower the mass of the galaxy. Old
ages for cluster ellipticals were similarly found by
\citet{2006ApJ...637..200Y}.

Estimates of the current to mean star formation rates, or
`birthrate,' of galaxies using population synthesis models for
galaxies of a range in mass and type show a trend of decreasing
birthrate with increasing stellar mass \citep{2001AJ....121..753B,
2002A&A...396..449G, kauffmann-2003-341, 2004MNRAS.351.1151B,
gallazzi-2005-362}, again indicating that the more massive a galaxy
is, the earlier most of its stars were formed.

The confluence of wide-field surveys and high-speed computing is now
making feasible the testing of galaxy formation scenarios by detailed
population synthesis models with a spectral resolution and coverage
previously impossible on a large scale. In this paper, we apply
population synthesis models to estimate the mass, metallicites, and
star formation histories of Luminous Red Galaxies (LRGs) drawn from the
Sloan Digital Sky Survey (SDSS) \citep{2000AJ....120.1579Y}. We have
several reasons for concentrating on LRGs:\ 1.\ they form a large
sample of galaxies chosen according to well-defined selection criteria
forming a well-defined sample \citep{eisenstein-2001-122}, 2.\ the
galaxies form a uniform population both photometrically
\citep{1973A&AS...10..201F,1977ApJ...216..214V}, and spectroscopically
\citep{1973A&AS...10..201F, 2003ApJ...585..694E}, 3.\ the galaxies are
bright, providing high signal-to-noise ratio spectra, 4.\ they
represent the extreme high mass end of galaxy stellar masses and an
extreme in anti-hierarchical behaviour, 5.\ star formation has ceased
in all but a few percent of the systems \citep{Rose06},
and so the influence of emission lines associated with
ongoing star formation on the key absorption features used to infer
the properties of the galaxies is minimal, 6.\ they are an
intrinsically interesting class of objects, being the most massive and
most strongly clustered galaxies in the Universe.

Population synthesis analyses for galaxies in SDSS Data Release 2
(DR2) and earlier data were performed previously by \cite{kauffmann-2003-341}
(hereafter K03), \cite{2004MNRAS.351.1151B} and \cite{gallazzi-2005-362} (hereafter G05).
These studies provide useful comparisons for our results. Our
analysis, however, differs in several respects from these earlier
efforts. We exploit high resolution synthetic spectra to model several
absorption indices that match the resolution of the spectra. The
analyses of K03 and \cite{2004MNRAS.351.1151B} used the 4000-\AA\
break ($D_n4000$) and Balmer absorption, or emission lines, to estimate
the star formation rates and histories. The closest population
synthesis analysis to ours is that of G05, who similarly based their
model-fitting on several high resolution absorption features. This
work, however, differs from theirs as follows:\ 1. an alternative
population synthesis model is used, utilising an independent set of
stellar templates to construct the spectra and a different set of
stellar evolution tracks, 2. the sample of LRGs analysed is
substantially larger, and 3. we use a different set of absorption
features specifically adapted to the LRG sample analysed rather than
the full galaxy population analysed by G05. In consequence, our
results are complementary to theirs, and provide independent checks on
their findings, while at the same time allowing us to extend the
results using the larger sample. G05 also found some discrepant
absorption indices poorly matched by the best-fitting models. Our
alternative set of models sheds light on the origin of these
discrepancies.

Another goal is to quantify the metallicities of the systems and their
deviations from solar abundances. Previous studies have suggested that
giant ellipticals and LRGs contain super-solar metallicities
\citep{1973A&AS...10..201F, 1985ApJ...296..340P, gallazzi-2005-362},
with indications of $\alpha$-enhanced abundances
\citep{1992ApJ...398...69W, 1993MNRAS.262..650D, trager-2000-119,
2000AJ....120..165T, 2003ApJ...585..694E}, although agreement is not
universal \citep{kelson-2006-642}. The galaxies also show a trend of
increasing metallicity with velocity dispersion or mass
\citep{2000AJ....120..165T, caldwell-2003-125,thomas-2005-621,
kelson-2006-642}. Little attention has been devoted to the modelling
of the metallicity distribution of early type galaxies. In
Section~\ref{sec:results}, we present models incorporating into standard
closed box and accreting box models the sudden quenching of
star-formation, as may occur if Active Galactic Nuclei heat the gas,
effectively shutting down gas cooling, to describe the evolution of
metallicity. We find that the measured metallicity distribution
provides added support to the star-formation quenching scenario in the
context of these enrichment models.

In the next section, we discuss the sample selected. In Section 3, we
describe our methods. Results from the analysis are presented in
Section 4. A discussion of the results and conclusions are provided in
Section 5.

\section{Sample selection}
\label{sec:SDSS}

The spectra used in this analysis are taken from the Sloan Digital Sky
Survey Data Release 4 (DR4) \citep{SDSSDR4-2006-162}. The SDSS is a
five-band imaging survey of the northern Galactic gap, combined with
an extensive spectroscopic follow-up program. The spectra cover a
large wavelength range (3800--9200~\AA) at high resolution
$\lambda/\Delta\lambda \sim 1800$ taken with 3 arcsec diameter fibres.
The target spectra are selected using the Luminous Red Galaxy (LRG)
cuts as defined in \citet{eisenstein-2001-122}. The aim of the SDSS
LRG survey was to produce a `volume-limited' sample of intrinsically
luminous ($L \geq 3L^{\star}$), intrinsically red galaxies out to
redshifts of $z=0.5$. Such a sample is useful as a probe of large
scale structure in the Universe. Here we use the data to study the
properties of giant elliptical galaxies.

The LRGs are selected from the SDSS database by requiring the
\verb#TARGET_GALAXY_RED# flag to be set. An additional cut requires
the redshift to fall within the range {$0.15 < z < 0.4$}, with the
lower limit in place to ensure that underluminous galaxies do not
enter the sample. A velocity dispersion cut is also made, requiring
{$70~$kms$^{-1} < \sigma < 450~$kms$^{-1}$}. This initial selection
yields a dataset of 24,615 galaxies.

Before detailed analysis was undertaken, each spectrum was blueshifted
to its respective rest-frame and resampled in steps of 2~\AA\ to match
the resolution of the model spectra. Unlike in DR1, the DR4 spectra
are not pre-corrected for foreground Galactic dust extinction. We applied
this correction using the maps of \citet{schlegel-1998-500}, combined
with an extinction law of \citet{cardelli-1989-345} and $R_V =
3.1$. All magnitudes used are also first corrected for foreground
extinction, using values given in the SDSS database.

Our analysis is based predominantly on fits to five specific
absorption indices, defined below in Section~\ref{subsec:diag}. LRG
spectra from the SDSS, however, contain not only absorption features
but also emission lines produced by hot, interstellar gas. These lines
can contaminate the stellar absorption lines and hence suppress the
measured absorption index values. Unlike \gal{}, who estimate the
effect of emission lines by fitting a linear combination of template
spectra, we choose to simply exclude those galaxies where emission
lines are detected within the index passbands. This is done by masking
pixels from the index calculation for which the \verb#SP_MASK_EMLINE#\
flag is set. We also reject pixels which have no data present, no sky
information, or bright sky levels (flags \verb#SP_MASK_NODATA#,
\verb#SP_MASK_NOSKY#\ and \verb#SP_MASK_BRIGHTSKY#). After this, an
index will only have a measured value if there is more than one
unmasked pixel in each of the three defining passbands. To make the
final cut, spectra must then have measured values for all five of the
matching indices described in Section~\ref{subsec:diag}. These cuts
reduce our sample to 4391 spectra in total. Selecting this subset of
galaxies does not systematically affect our results; when parameter
estimates are made using a subset of the five fitting indices, the
overall distribution of that parameter is the same as when all five
indices are present. This suggests our findings for the LRG population
as a whole are not much affected by contamination from galaxies with
appreciable emission lines. This is expected from the low (few per
cent) contribution of actively star-forming galaxies to the LRG
population \citep{Rose06}.

\section{Method}
\label{sec:method}

\subsection{P\'EGASE synthetic spectra}
\label{subsec:Pegase}

To analyse the physical properties of the LRG sample, we use the
P\'EGASE evolutionary synthesis code \citep{PEGASE-1997-326}, version
2.0, coupled with a new library of high resolution Kurucz stellar
spectra generated by \citet{murphy-2004-351}. With this version of the
code, developed by \citet{murphy-2003-thesis}, we can produce spectra
of stellar populations with a resolution of 2~\AA\ in the wavelength
range 3000--10000~\AA, ideal for comparing with the SDSS spectra. The
exceptions are for a subset of cool stars, for which only
20~\AA\ resolution synthetic spectra are available, and similarly for
very hot stars, although these tend to be nearly featureless so that
high resolution is not crucial. Details of the synthetic spectra used
are provided by \citet{murphy-2004-351} and
\citet{murphy-2003-thesis}.

The P\'EGASE code uses the standard PADOVA stellar evolutionary tracks
of \citet{bressan-1993-100}, supplemented by those of
\citet{1997ApJ...477..313A} for low mass stars becoming helium white
dwarfs. Pseudo-tracks of thermally pulsing asymptotic giant branch
(AGB) stars are constructed following
\citet{1993A&A...267..410G}. Finally, post-AGB and CO white dwarf
tracks are taken from \citet{1995A&A...299..755B} and
\citet{1983ApJ...272..708S}. The positions in the HR diagram of
unevolved low-mass stars are taken from \citet{1997A&A...327.1039C}.
\citet{bruzual-2003-344} find that the P\'EGASE predictions for
colours and mass-to-light ratios agree closely with those of their
standard population synthesis model for simple stellar populations
(SSPs) older than 1~Gyr when the same IMF is adopted.

We note that the evolutionary tracks are computed with solar abundance
ratios, while our galaxies are expected to have ratios that differ
from the solar values. This will dictate the choice of stellar
diagnostics used to constrain LRG properties, and is discussed further
in the next section.

We adopt a Salpeter initial mass function (IMF) for masses
$0.1-120\,M_\odot$. This choice differs from those of \kau{} and
\gal{}, who adopt the \citet{2001MNRAS.322..231K} and
\citet{2003PASP..115..763C} IMFs, respectively. (Several IMFs are
provided by Chabrier; the form adopted by Gallazzi et al. was based on
{\it Hipparcos} measurements of disc stars in the Milky Way.)  Both
groups perform their modelling using the Bruzual--Charlot
\citep{bruzual-2003-344} population synthesis code. Tests reported in
\citet{bruzual-2003-344} on simple stellar populations show that the
Salpeter IMF predicts a mass-to-light ratio a factor 1.5--2 greater
than these other choices, which flatten at the low mass end compared
with the Salpeter IMF. For masses above a solar mass, both IMFs are
very similar in shape to the Salpeter IMF. Since the light of the LRGs
is dominated by more massive stars near the Main Sequence turn-off, we
expect that the inferred properties of ages and metallicities will be
largely unaffected by the choice of IMF. This is supported by the
similar colours produced by the different IMFs
\citep{bruzual-2003-344}. Since there are no direct measurements of
the IMF in LRGs, we adopt the Salpeter IMF for the sake of its
widespread usage.

We also differ from \kau{} and \gal{} in the stellar spectral
libraries used. While we rely on synthesised stellar spectra to
provide near uniform coverage of the space in surface temperature,
surface gravity and metallicity, the Bruzual--Charlot code
incorporates the measured spectra of stars from the STELIB stellar
spectral library \citep{2003A&A...402..433L}. The STELIB library has
the advantage of having been drawn from actual stellar spectra, but at
somewhat lower resolution ($\sim3$\AA\ FWHM). The coverage in
metallicity, surface temperature and surface gravity is also not
uniform \citep{bruzual-2003-344, murphy-2003-thesis}. Our results may
be viewed as complementary to these earlier efforts, in that they test
the robustness of the inferred properties of the galaxies to the
assumptions made in the population synthesis modelling. We compare our
results with those of \kau{} and \gal{} and comment on any differences
in Section~\ref{sec:discussion}. Here we note that restricting the
modelling parameters and galaxy sample to those of \kau{} and \gal{}
results in good agreement with their findings.

\subsection{Stellar diagnostics}
\label{subsec:diag}

In this study we use absorption indices as stellar diagnostics to
model the strength of absorption features in the LRG spectra. The
original set of 21 Lick indices is considered
\citep{burstein-1984-287, worthey-1994-94}, augmented with the four
higher-order Balmer line indices of \citet{worthey-1997-111} and the
high-resolution H$\gamma_{HR}$ index of \citet{jones-1995-446}.  Each
index is measured by calculating the equivalent width over a central
bandpass, normalised by a pseudo-continuum which itself is defined by
two adjacent bandpasses. Previous studies have shown that some indices
are sensitive mainly to age, such as the Balmer lines, while others
are primarily sensitive to the metallicity of the stellar population,
such as the Fe and Mg indices \citep{trager-2000-119}.  Errors on the
indices are calculated following the method of
\cite{cardiel-1998-127}.

In addition, we consider the 4000~\AA\ break index D(4000), which gives
the ratio of the average flux density in the bands 4050--4250~\AA\ and
3750--3950~\AA\ \citep{bruzual-2003-344}. Also calculated is the
narrow-band version D$_n(4000)$ of \citet{balogh-1999-527}
(4000--4100~\AA\ and 3850--3950~\AA), which is believed to be
considerably less sensitive to reddening effects. The ratio of the
strength of the CaII H and K lines, as integrated from 3921--3946~\AA\
and 3956--3981~\AA, is also considered briefly
\citep{SDSSEDR-2002-123}. Unlike the original Lick index studies,
which use analytical fitting functions \citep{worthey-1994-94} to
model index trends, we follow the same method as \gal{}, directly
comparing indices measured from our high resolution synthetic spectra
to those from the SDSS.

The stellar spectra used are all based on solar metal abundance
ratios. External galaxies, as well as having a range of metallicities,
are expected to have varying abundance ratios of $\alpha$ elements to
Fe-peak elements. It is believed that $\alpha$ elements, which include
N, O, Mg, Ca, Na, Ne, S, Si and Ti, are produced predominantly in Type
II supernovae, while Fe-peak elements, which include Cr, Mn, Fe, Co,
Ni, Cu and Zn, originate mainly in Type Ia supernovae. If this
abundance ratio, denoted $\alpha/{\rm Fe}$, differs from the solar
value, then the models may show discrepancies with the data. A remedy
for this problem suggested by \citet{thomas-2003-339} is to combine Mg
and Fe indices in varying proportions to give a composite index that
is dependent on metallicity, but insensitive to $\alpha$ enhancement.

In this study we consider the $[\textrm{Mg}\textrm{Fe}]'$ index
defined by \citet{thomas-2003-339},
\begin{equation}
[\textrm{Mg}\textrm{Fe}]'=\sqrt{\textrm{Mg}_b(0.72\textrm{Fe5270}+0.28\textrm{Fe5335})}
\end{equation}
and the two composite indices of \citet{bruzual-2003-344},
\begin{equation}
[\textrm{Mg}_1\textrm{Fe}]=0.6\textrm{Mg}_1+0.4\log(\textrm{Fe4531}+\textrm{Fe5015})
\end{equation}
\begin{equation}
[\textrm{Mg}_2\textrm{Fe}]=0.6\textrm{Mg}_2+0.4\log(\textrm{Fe4531}+\textrm{Fe5015})
\end{equation}
The latter two indices were designed to provide a good match to
\citet{bruzual-2003-344} spectra for a broad range of galaxy types,
but initial studies here showed that they do not give as good a
fit to the LRG spectra compared with our P\'EGASE models. For this
reason we propose two additional composite indices, designed to be
independent of the abundance ratio, whilst also providing
observational values consistent with those measured for the LRGs using
the P\'EGASE model spectra:
\begin{equation}
[\textrm{Mg}_1\textrm{Fe}]'=0.8\textrm{Mg}_1+0.2\log(\textrm{Fe5709}+\textrm{Fe5782})
\end{equation}
\begin{equation}
[\textrm{Mg}_2\textrm{Fe}]'=0.8\textrm{Mg}_2+0.2\log(\textrm{Fe4383}+\textrm{Fe5406})
\end{equation}

\begin{figure}
\includegraphics[width=8cm]{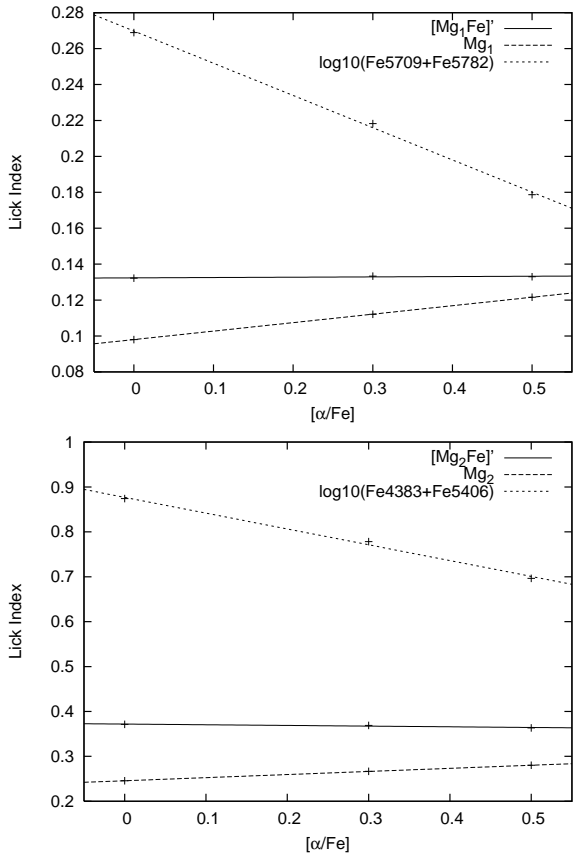}
\caption{Absorption indices as a function of $\alpha$/Fe at fixed overall
metallicity. Shown are the new $[\textrm{Mg}_1\textrm{Fe}]'$ (top)
and $[\textrm{Mg}_2\textrm{Fe}]'$ (bottom) indices with their
constituent Mg and Fe components. Data for the plots comes from
\citet{thomas-2003-339} and is representative of a 12~Gyr old SSP
model with solar metallicity (cf. figure~7 of
\citet{thomas-2003-339}).}
\label{plot:mgfe}
\end{figure}

Using data from \citet{thomas-2003-339}, Figure~\ref{plot:mgfe} shows
that the two indices are, in fact, insensitive to changes in $\alpha$
enhancement.\footnote{The data used in the plot are available
electronically at {www-astro.physics.ox.ac.uk/\~{}dthomas}} For both
indices, this behaviour is almost independent of age or metallicity.
In the case of $[\textrm{Mg}_1\textrm{Fe}]'$, $\alpha$/Fe
insensitivity holds for models older than $\sim1$~Gyr, and provides a
good metallicity indicator.  The $[\textrm{Mg}_2\textrm{Fe}]'$ index
is less well behaved, with $\alpha$ enhancement becoming apparent for
models younger than $\sim4$~Gyr, especially for low metallicity
populations.

To successfully extract age and metallicity information from the SDSS
spectra, we choose to simultaneously fit age-sensitive and
metal-sensitive indices that are both well reproduced by our models
and that show a weak $\alpha$/Fe dependence. After extensive studies
fitting a variety of index combinations from P\'EGASE to those from
the LRG sample, we settle on the following set of indices to be used
in the final analysis. The indices $[\textrm{Mg}\textrm{Fe}]'$ and
$[\textrm{Mg}_1\textrm{Fe}]'$ are chosen for their sensitivity to
metallicity. It was decided that $[\textrm{Mg}_2\textrm{Fe}]'$ should
not be used as it is too dependent on $\alpha$/Fe and does not provide
an adequate fit to our models. For age sensitive indices we choose
H$\beta$, H$\gamma_F$ and D$_n(4000)$. Despite the fact that
H$\gamma_F$ has been shown to be sensitive to element abundance ratios
\citep{thomas-2003-339,korn-2005-438}, including it does not
systematically affect our results.

\subsection{Star formation library}
\label{subsec:sflib}

We adopt a Bayesian statistical approach to derive the star formation
histories, metallicities and masses for each galaxy in our sample. The
method is similar to the one described in \kau{} and \gal{}, where the
data are assumed to be drawn from a distribution of models, each
described by a parameter vector $\vec{P}$. A likelihood distribution
of a given parameter may be obtained by comparing the observational
data with a set of models that populate the space of all possible
$\vec{P}$. This set of models follows a prior distribution that
represents any knowledge of various $\vec{P}$ values in the absence of
any data.

As in \kau{}, we generate a library of 31000 Monte Carlo realizations
of different star formation histories. Our models are characterised by
an underlying model where stars have been forming according to the
exponential law ${\rm SFR}\propto\exp(-\gamma t)$ since time $t_{\rm
form}$. Random bursts are then superimposed on this continuous model,
and can occur at any time after $t_{\rm form}$ with equal probability.

Some of our parameter choices differ from those of \kau{} and \gal{}
both because our analysis is restricted to LRGs rather than the broad
range of galaxy types considered in these earlier studies, and to
allow a wider possible range in the prior. We take $t_{\rm form}$ to
be uniformly distributed between 13.5 and 1.5~Gyr and $\gamma$ over
the interval 0 to 2. This differs from \gal{}, who take $\gamma$
between 0 and 1. We also set the fraction of models that have
experienced a burst in the last 2~Gyr to 1 per cent, compared with 20
per cent used by \kau{} and 10 per cent used by \gal{}. The
implications of changing these priors is dicussed in
Section~\ref{subsec:sfh}. During a burst, stars can form at a constant
rate for a time $t_{\rm burst}$, uniformly distributed from $3\times
10^7$ to $3\times 10^8$~yr. The mass of stars formed in a burst,
$M_{\rm burst}$, is characterised by $A=M_{\rm burst}/M_{\rm cont}$,
where $M_{\rm cont}$ is the total mass of stars formed by the
continuous model. $A$ is distributed logarithmically in the range 0.03
to 4.0. The metallicity of the models is logarithmically distributed
from $0.02-2.5Z_{\sun}$ and is permitted to take on different values
for the continuous model and burst. In most cases, however, for
simplicity we shall quote the overall light-weighted metallicity of
the model.

Multiple versions of each model are then constructed by gaussian
smoothing the spectra to mimic the stellar velocity dispersion of a
galaxy. In the range $150 < \sigma < 300~\textrm{kms}^{-1}$, spectra
are smoothed at 10~kms$^{-1}$ intervals, rising to 20~kms$^{-1}$ for
$300 < \sigma < 450~\textrm{kms}^{-1}$. These intervals are chosen so
that over 95 per cent of LRG spectra have a velocity dispersion
measurement that lies within one standard deviation of a model
value. (The mean velocity dispersion error for the sample is always
larger than 20~kms$^{-1}$.) This gives us a total of 23 libraries,
each smoothed with a different velocity dispersion and each containing
31000 star formation histories.

For each spectrum in these libraries, we compute the strengths of the
H$\beta$, H$\gamma_F$, D$_n(4000)$, $[\textrm{Mg}\textrm{Fe}]'$ and
$[\textrm{Mg}_1\textrm{Fe}]'$ indices. The apparent $u$, $g$, $r$, $i$
and $z$ magnitudes are calculated using SDSS filters at redshifts
between 0.15 and 0.4 in intervals of 0.01. Unless otherwise stated, we
assume a flat Friedmann-Robertson-Walker cosmology, with $\Omega_m =
0.25$, $\Omega_{\Lambda} = 0.75$ and $H_0 =
73~\textrm{kms}^{-1}\textrm{Mpc}^{-1}$.

\subsection{Parameter estimation}
\label{subsec:pe}

The probability density functions (PDFs) of various physical
parameters are calculated by comparing our set of index strengths for
a single galaxy to every model spectrum in the library. First of all
we select the library with the velocity dispersion closest to the
observed value. Each model is assigned a weight $w=\exp(-\chi^2/2)$,
where $\chi^2$ is calculated as follows:
\begin{equation}
\chi^2 = \sum_{i=1}^{i=5}{\left(\frac{I_i^{\rm mod}-I_i^{\rm obs}}{\sigma_I}\right)^2},
\end{equation}
where $I_i^{\rm mod}$, $I_i^{\rm obs}$ and $\sigma_I$ are the model
index, observed index and error values respectively. The sum runs over
all five indices in our set. The PDF of a particular parameter is then
given by the distributions of the weights $w$ of all the models in the
library. We characterize these distributions using the mode (the peak
of the distribution) and the median. Errors are often quoted using the
$16-84$ percentile range of the PDF, corresponding to $\pm1\sigma$ for
a gaussian distribution.

\begin{figure*}
\begin{center}
\begin{tabular}{cc}
   \includegraphics[angle=-90,width=8cm]{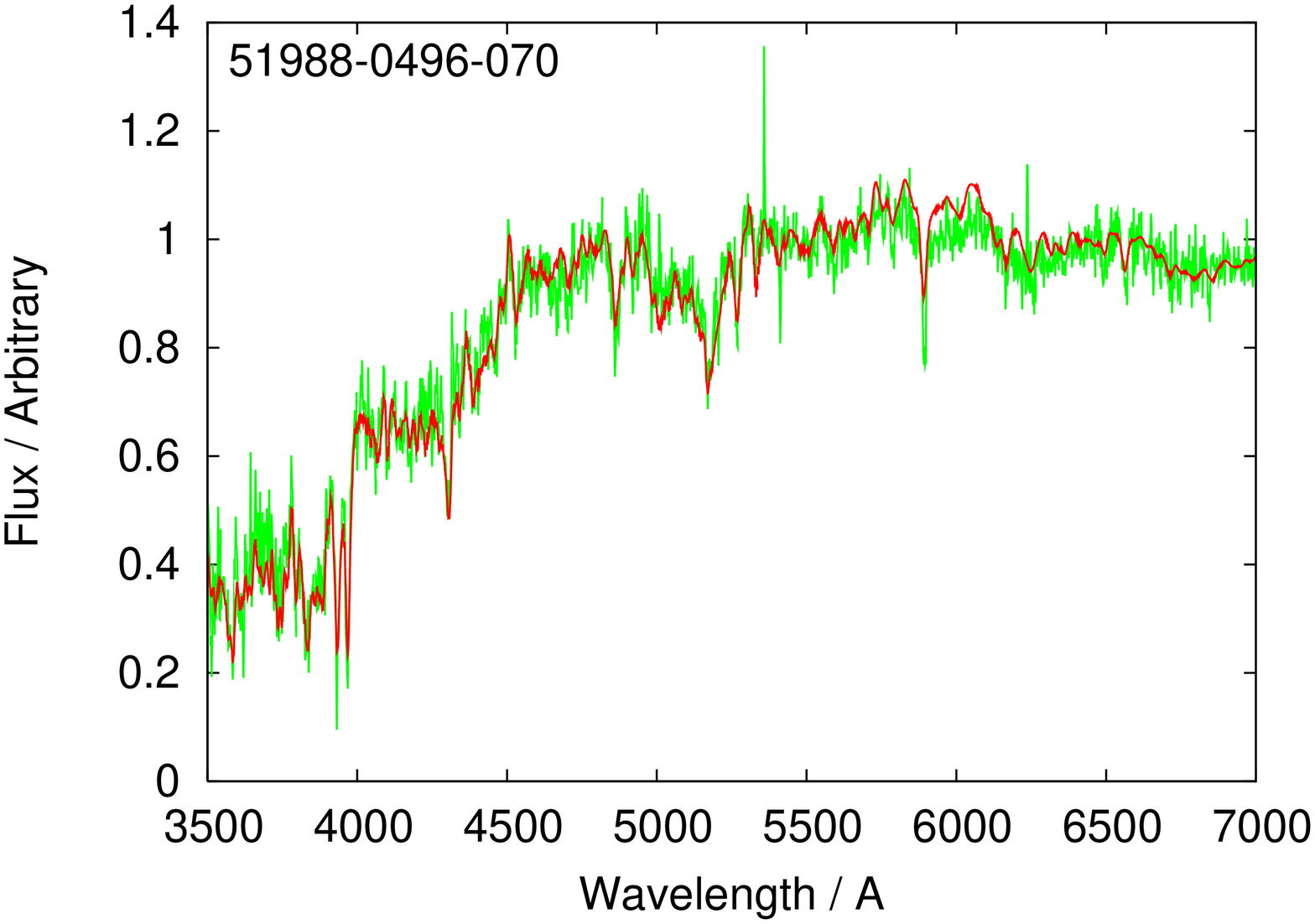} &
   \includegraphics[angle=-90,width=8cm]{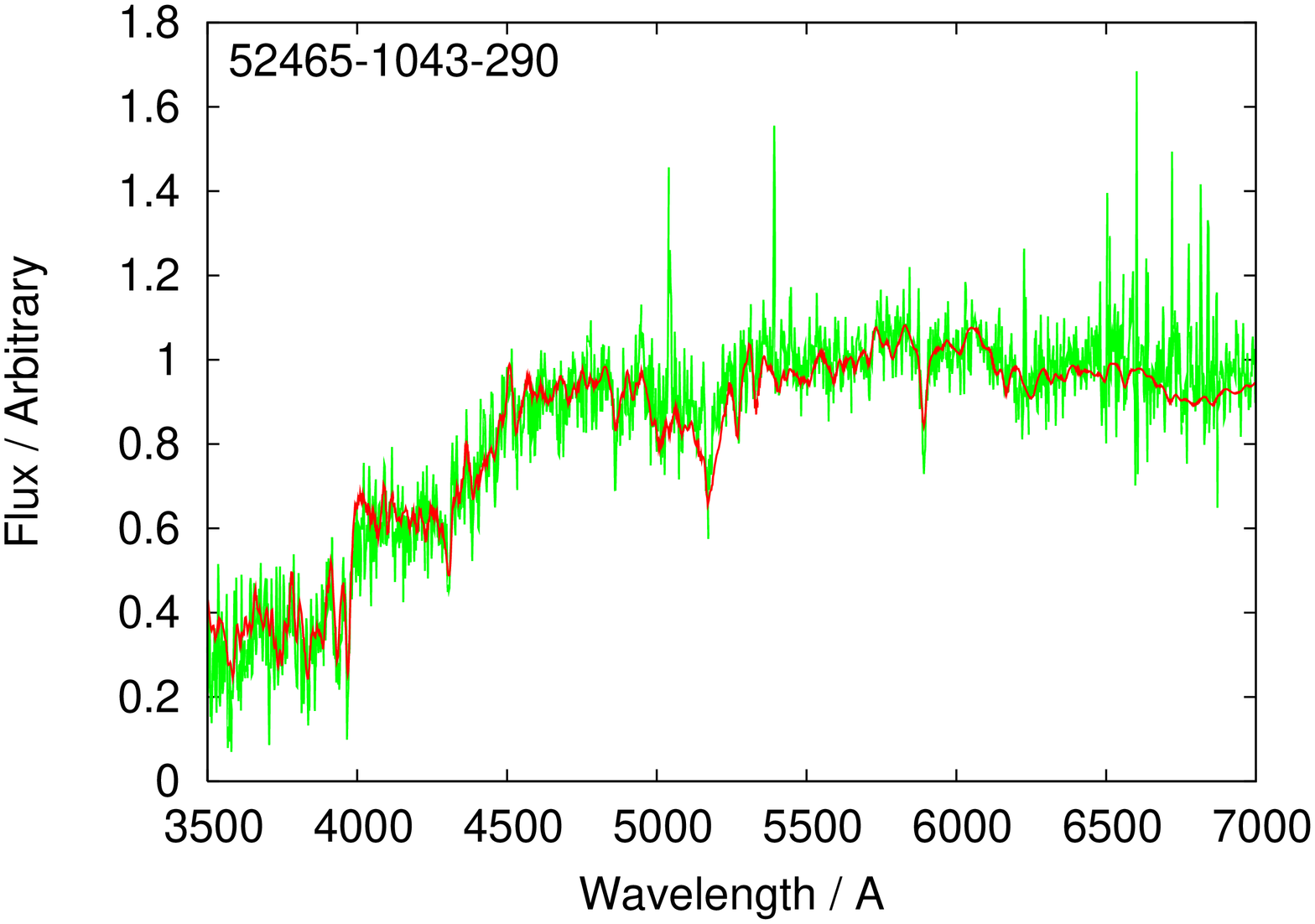} \\
   \includegraphics[angle=-90,width=8cm]{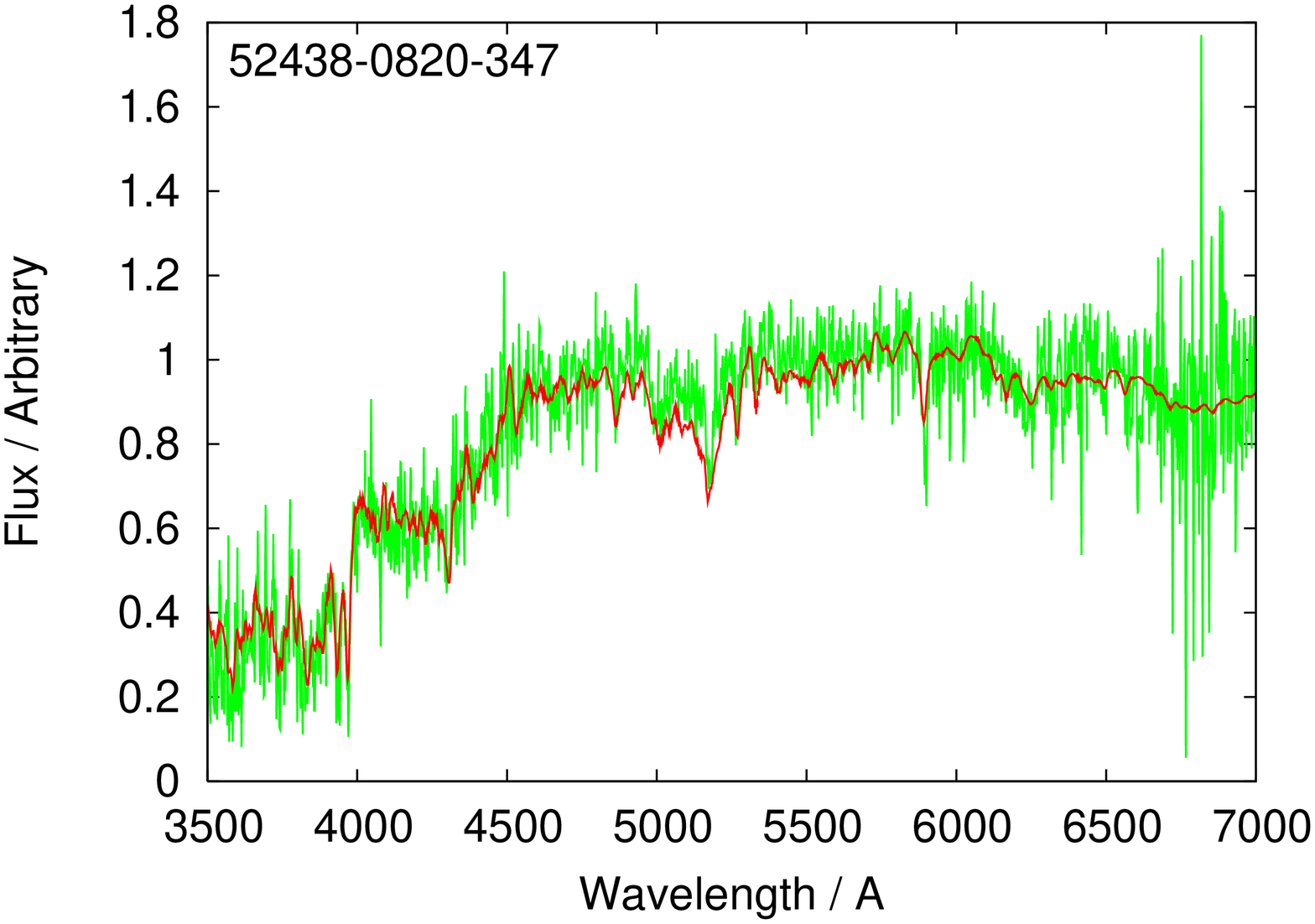} &
   \includegraphics[angle=-90,width=8cm]{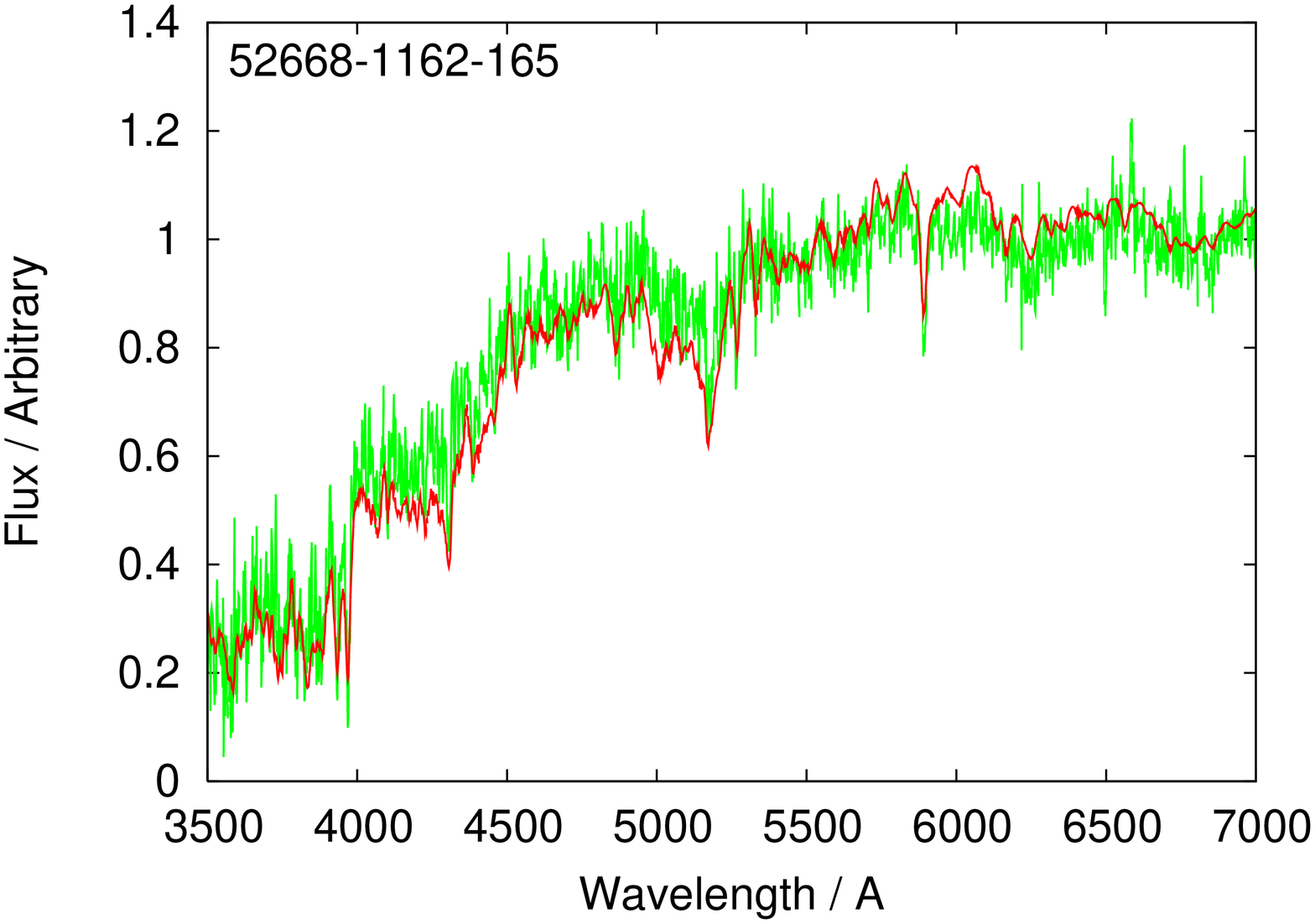} 
\end{tabular}
\caption{Examples of LRG spectra from the SDSS are shown in green, after
dereddening for Milky Way extinction.
The P\'EGASE spectra from the Monte Carlo library with the best-fitting
set of the selected five absorption indices (see text) are overplotted in red.
Top-Left: The best fit shows a young population with a light-weighted age of
3.5~Gyr and metallicity $Z = 0.03$.
Top-Right: An example of an old population that formed
9.9~Gyr ago, which experienced a significant burst 1.7~Gyr
ago. The best-fit metallicity is $Z=0.02$.
Bottom-Left: An example of a spectrum that our method finds a good fit for
despite a low SNR (6.3). The galaxy is 5.7~Gyr old with a metallicity
of $Z = 0.03$.
Bottom-Right: An example of a spectrum for which
our library fails to find a good fit. The best fitting set of
indices gives $\chi^2 = 19.7$ for five degrees of freedom. The
previous three spectra have $\chi^2 = 4.7,5.2,2.0$ respectively. The
model spectra have been smoothed slightly for presentation.}
\label{fig:spectra}

\end{center}
\end{figure*}

In Figure~\ref{fig:spectra}, we show some examples of spectra from the
LRG sample, with the model spectrum that gives the best fit based on
our set of five indices. In general the best fitting spectrum
reproduces many of the absoption features correctly. However, as we
only fit absorption indices, the overall shape of the continuum is less
well reproduced, especially towards the red end of the spectrum. The
strength of this method is that we can derive the star formation
history independently of the shape of the continuum and the extinction
model. We only need an estimate of the dust extinction when computing
mass estimates.

To characterize the age of each galaxy in our library, we calculate
the bolometric light-weighted age
\begin{equation}
t_L = \frac{\int_0^t{SFR(t-\tau) L(\tau) \tau d\tau}}
{\int_0^t{SFR(t-\tau) L(\tau) d\tau}},
\end{equation}
where $L(\tau)$ is the bolometric luminosity of the system evolved
over a time $\tau$. Similarly, we can define a light-weighted
metallicity
\begin{equation}
Z_L = \frac{\int_0^t{SFR(t-\tau) L(\tau) Z(t-\tau) d\tau}}
{\int_0^t{SFR(t-\tau) L(\tau) d\tau}},
\end{equation}
where $Z(t-\tau)$ is the metallicity of the stars at the time
$t-\tau$. The metallicity may also be written as $[Z/H] =
\log(Z/Z_{\sun})$, where $Z_{\sun} = 0.02$. In this study we will
refer to these quantities as the age and metallicity of a model,
respectively.

\begin{figure}
\includegraphics[angle=-90,width=8cm]{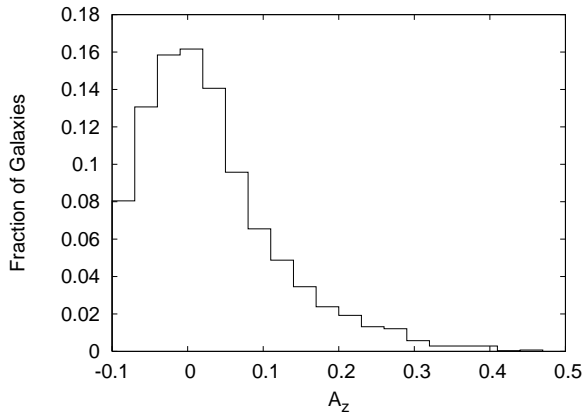}
\caption{The distribution of dust attenuation values in the $z$ band
for the 4391 LRGs in our sample.}
\label{plot:azdist}
\end{figure}

To estimate the stellar mass of a galaxy, we first need to calculate
the effect of dust on the spectrum. Here we model the dust as a simple
slab of material between us and the galaxy. Any light, $S(\lambda)$,
from the galaxy is attenuated according to the following
wavelength-dependent law,
\begin{equation}
S_{\rm obs}(\lambda) = S_{\rm emit}(\lambda)\times 10^{-0.4\times A_{\lambda}}
\end{equation}
where $A_(\lambda)$ is given by the extinction curve of
\citet{cardelli-1989-345}. This empirical model takes the following
two-component form:
\begin{equation}
\frac{A_{\lambda}}{A_V} = a(\lambda) + \frac{b(\lambda)}{R_V}
\label{eq:cardelli}
\end{equation}
where $a(\lambda)$ and $b(\lambda)$ are pre-determined polynomials. We
adopt the standard Milky Way value of $R_V=3.1$ when fitting dust
models to the data. This gives essentially a single-component dust
model, characterised by $A_V$, the attenuation in the V-band (centred
on 5500~\AA). We infer the value of $A_V$ using the difference between
the fibre $r-i$ colour and the $r-i$ colour of each model at a
redshift matching the data,
\begin{equation}
A_V = \frac{(r-i)_{\rm fibre} - (r-i)_{\rm model}}{\left[a(r) +
\frac{b(r)}{R_V}\right]-\left[a(i) + \frac{b(i)}{R_V}\right]}.
\end{equation}
Using equation~\ref{eq:cardelli}, we then calculate the extinction-corrected,
photometric $z$-band luminosity of the entire galaxy. The stellar mass for a
given model can be estimated as follows,
\begin{equation}
M_* = L_z \times \left(\frac{M_*}{L_z}\right)_{\rm model},
\end{equation}
where $(M_*/L_z)_{\rm model}$ is the redshifted mass-to-light ratio in the
$z$-band. This assumes that the mass-to-light ratio within the fibre
is the same as for the overall galaxy. G05 find no significant trend
of age or metallicity with the fraction of total light entering the
fibre for massive early-type galaxies, so this seems a fairly safe
assumption. The mass values for each model are then weighted as usual
to give a PDF for the stellar mass. As in \gal{}, we allow
attenuations down to $A_z = -0.1$ to account for errors in
measurements and discrepancies within the models.
Figure~\ref{plot:azdist} shows the distribution of median $A_z$ values
for our sample of SDSS LRGs. The typical attenuation in the $z$ band
is quite small, with a median value of $2.6 \times 10^{-3}$. This is
consistent with the picture of LRGs as early-type galaxies containing
old stellar populations with little dust and gas.

\section{Results}
\label{sec:results}
We now apply the above analysis to the SDSS sample of 4391 Luminous
Red Galaxies as described in Section~\ref{sec:SDSS}.

\subsection{Index fitting}
\label{subsec:fits}

\begin{figure*}
\includegraphics[width=15.5cm]{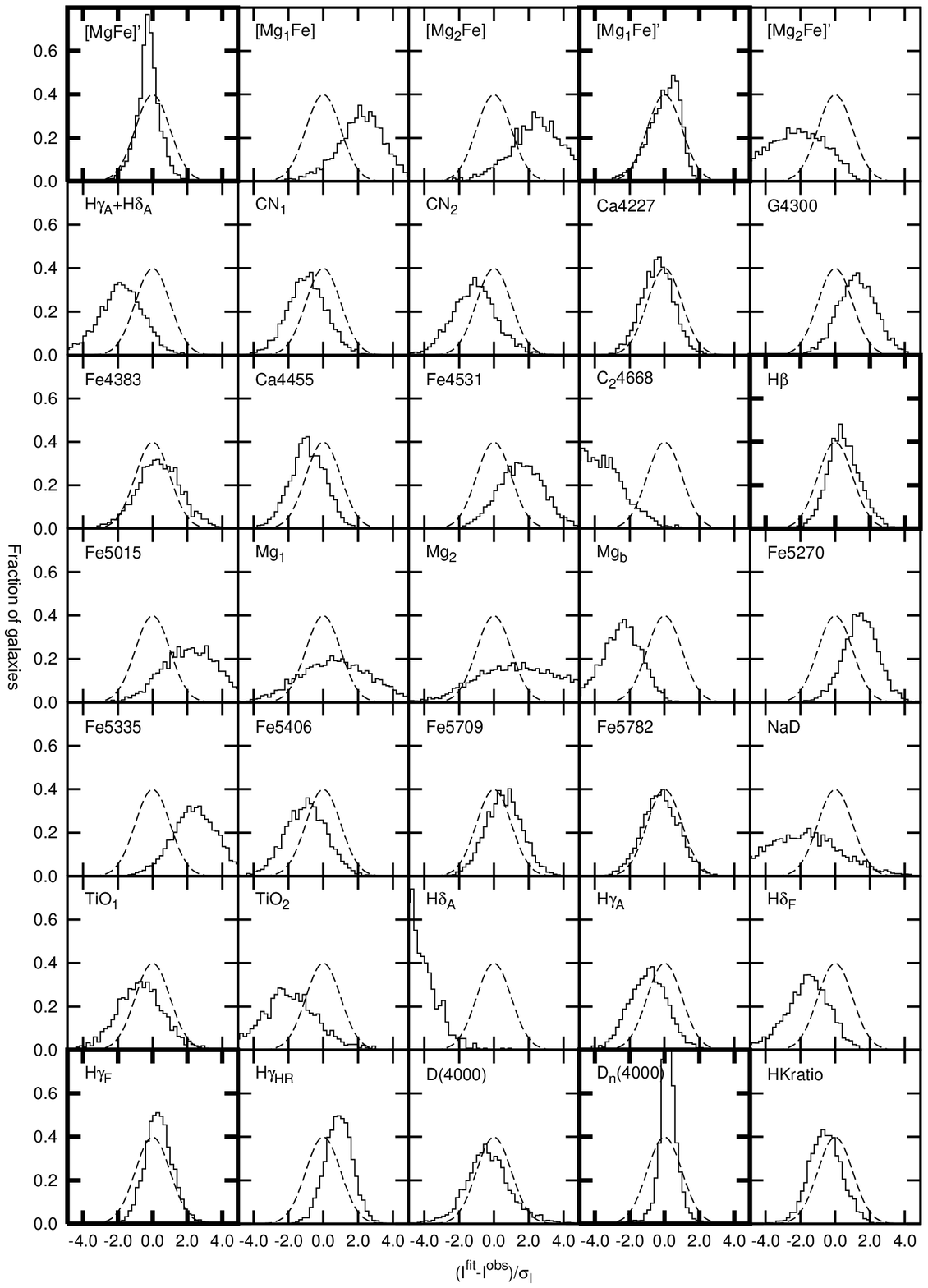}
\caption{Simultaneous fits of the H$\beta$, H$\gamma_F$, D$_n(4000)$,
$[\textrm{Mg}\textrm{Fe}]'$ and $[\textrm{Mg}_1\textrm{Fe}]'$ indices
(highlighted) of the 4391 galaxies from the SDSS LRG sample with
measurable indices. Each panel shows the distribution of the best
fitting index $I^{\rm fit}$ minus the observed one, $I^{\rm obs}$,
divided by the index error $\sigma_I$. The dashed line in each panel
indicates a gaussian with unit standard deviation.
\label{fig:lick}}
\end{figure*}

Figure~\ref{fig:lick} shows the distribution of indices in the LRG
sample when fitted with the final set of indices, using the method
described in Section~\ref{subsec:sflib} (cf., figure~18 of
\citet{bruzual-2003-344}). While this plot shows that many indices
agree very well with models (e.g. $[\textrm{Mg}\textrm{Fe}]'$,
$[\textrm{Mg}_1\textrm{Fe}]'$, Ca4227, Fe5782, H$\beta$, H$\gamma_F$,
D$_n(4000)$) others show complete failure (e.g. C$_2$4668, NaD,
H$\delta_A$). \citet{thomas-2003-339} and \citet{korn-2005-438}
investigate the effect of varying individual element abundances on the
Lick indices. \citet{thomas-2003-339} find that model predictions for
C$_2$4668 poorly reproduce globular cluster data, even when Mg$_1$,
CN$_1$ and CN$_2$ are well-modelled. \citet{korn-2005-438} find this
index is most sensitive to the carbon abundance. They also find that
the NaD index is sensitive to both sodium and metallicity and
H$\delta_A$ is affected by iron abundance and metallicity. The
narrower passband of H$\delta_F$ reduces the sensitivity to iron. The
blue indices CN$_1$, CN$_2$, Ca4227, G4300, Ca4455, Fe4531 and
C$_2$4668, along with TiO$_2$, are known to suffer the worst modelling
uncertainties \citep{korn-2005-438}. Many of the remaining
discrepancies are likely caused by overabundances of individual
elements that are not accounted for by the P\'EGASE model. For
instance, the iron index Fe5015 is sensitive to Ti and Mg in addition
to iron, and to $\alpha$ enhancement as well \citep{thomas-2003-339}.
\citet{thomas-2003-339} also find that the Mg$_b$ index is sensitive
to $\alpha$ enhancement, more so than are either the Mg$_1$ or Mg$_2$
indices, for which the agreement is better, though the spread is still
large. The underprediction of some indices, like $\textrm{CN}_1$ and
$\textrm{CN}_2$, is expected for $\alpha$-enhanced abundances
\citep{thomas-2003-339}. We discuss the indices further in
Section~\ref{sec:discussion}.

\subsection{LRG properties}
\label{subsec:properties}

\begin{figure}
\includegraphics[width=8cm]{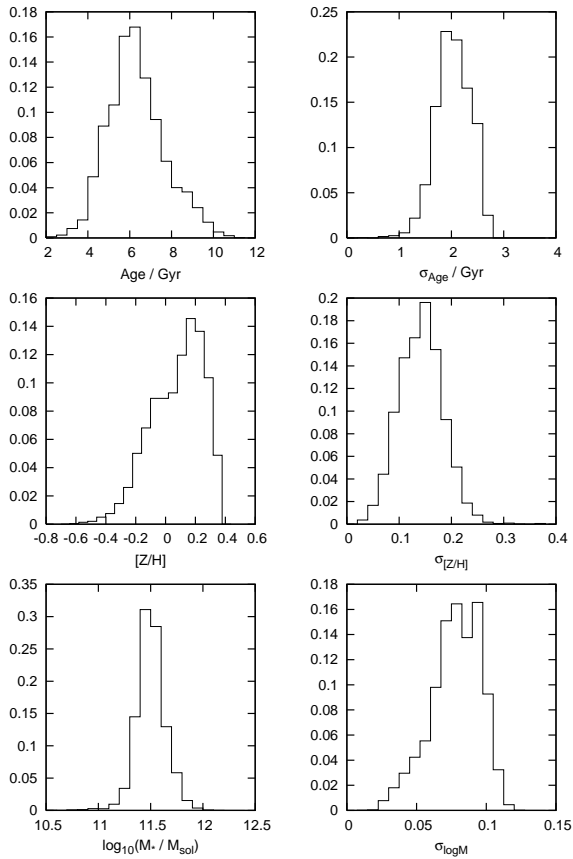}
\caption{Histograms of the median values of age (top-left), metallicity (middle-left)
and stellar mass (bottom-left) for 4391 Luminous Red Galaxies from the SDSS-DR4. The
y-axis units give the fraction of galaxies in a particular bin. The right-hand panels
show the respective error distributions, calculated as one-half of the 16--84
percentile range.\label{fig:agezmass}}
\end{figure}

Here we present the results obtained for the light-weighted age,
metallicity and stellar mass of the 4391 Luminous Red Galaxies in our
sample. Figure~\ref{fig:agezmass} shows the distributions of the
median-likelihood estimates of these parameters (left-hand side) and
their errors (right-hand side) in the top, middle and bottom panels
respectively. The age distribution shows a peak at around 6~Gyr,
suggesting the LRGs are comprised of mainly old stellar
populations. Typical uncertainties on the age are $\sigma_t \sim
2~\textrm{Gyr}$, with no galaxies showing errors larger the 3~Gyr.
The distributions also indicate an average super-solar metallicity of
$[Z/H] \sim 0.2$, with a long tail down to $[Z/H] \sim -0.6$. Errors
in the metallicity measured in individual galaxies are around
$\sigma_{[Z/H]} \sim 0.15$, extending up to 0.3 for some
galaxies. Similar distributions are found for the age and metallicity
of early type galaxies by \citet{trager-2000-119}.  Finally, the mass
shows a narrow distribution that peaks at $\log(M_*/M_{\sun}) \sim
11.5$. It is fairly well constrained, with a maximum error of
$\sigma_{\log(M_*/M_{\sun})} \sim 0.1$. The high mass of the sample is
expected from the LRG high luminosity cuts. The narrow distribution
suggests that we have indeed selected a homogeneous population of
galaxies.

\subsection{Age, mass and metallicity relationships}
\label{subsec:relationships}

We are also interested in exploring relationships between the mass,
age and metallicity of the galaxies in our sample. Rather than simply
plot the median of two separate distributions, we calculate the joint
distribution of each pair of parameters. This is done by adding the 2D
normalised probability distribution of all galaxies in our sample,
giving each galaxy equal weight.

In this study, we use the same cosmology as that of
\citet{delucia-2006-366}, who use merger trees from the Millenium
Simulation, grafted onto semi-analytic models of galaxy formation, to
track the properties of elliptical galaxies. As LRGs are early-type
galaxies, we can directly test the predictions of these simulations at
the high-mass end using the LRG sample.

\begin{figure*}
\includegraphics[width=15cm]{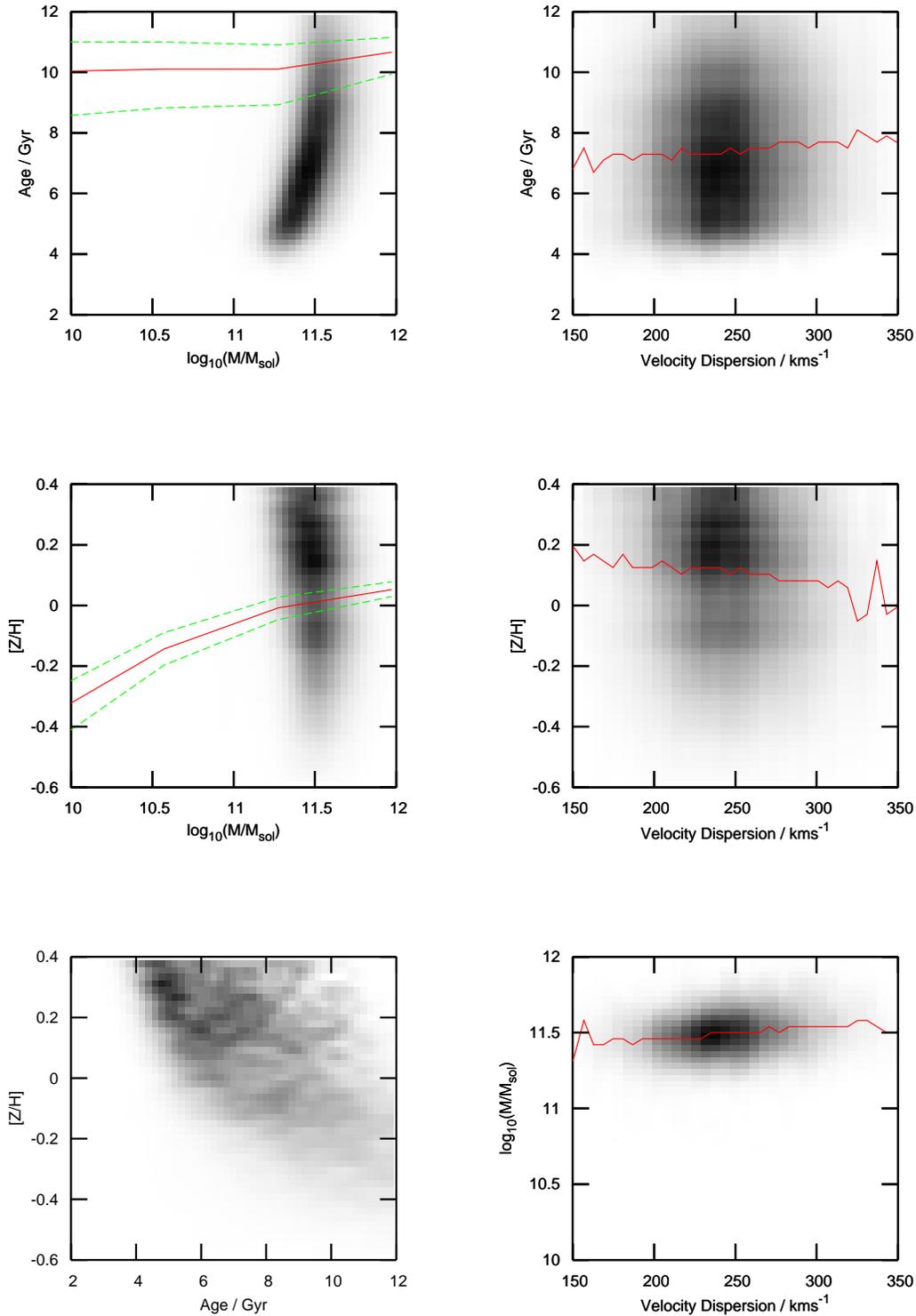}
\caption{Two-dimensional probability distributions for various pairs
of galaxy properties. The plots were obtained by adding the normalized
2D distributions for all 4391 galaxies in the SDSS-DR4 sample of LRGs.
The left hand-column shows light-weighted age (at $z=0$) and metallicity
against stellar mass, and metallicity against age, from top to bottom
respectively. The right-hand column shows age, metallicity and mass
against velocity dispersion.
The solid line in the upper two plots of the left-hand column
shows the median value as predicted by \citet{delucia-2006-366}, using
semi-empirical models. The dashed lines represent the upper and lower
quartiles of their distribution. The solid lines in the right-hand column
plots give the median parameter value in each velocity dispersion bin.
\label{fig:twofinal}}
\end{figure*}

In the top-left panel of Figure~\ref{fig:twofinal}, we show the 2D
probability distribution of age against mass. In this plot, the age
has been estimated for each galaxy at a redshift of $z = 0$, by adding
the lookback time to the light-weighted age. Also shown is the
distribution of metallicity against stellar mass in the middle-left.
These two plots are comparable to figure~6 of
\citet{delucia-2006-366}, whose predictions for the same variables are
shown as the solid and dashed lines, representing the median and upper
and lower quartiles respectively.

The plots do not show any strong trends with stellar mass, and reflect
the distributions seen in Figure~\ref{fig:agezmass}. The age appears
to be slightly correlated with mass, and suggests that the less
massive galaxies are made up of younger stellar populations within the
restricted mass range of the LRGs. The metallicity appears to be
fairly uniform over the stellar mass range studied, with at most a
slight anti-correlation with mass. The slight bimodality in the
distribution is discussed further in Section~\ref{subsec:z}. The ages
predicted by \citet{delucia-2006-366} exceed our estimates by
1--3~Gyr. Their metallicity predictions agree closely with ours,
although their predicted values are lower by about 0.1~dex.

An issue that invariably arises when deriving galaxy properties is the
age-metallicity degeneracy, which occurs due to competition between
age and metallicity on absorption-line strengths, so that old,
metal-poor galaxies can be mistaken for young, metal-rich ones (or
vice versa). Figure~\ref{fig:twofinal}, bottom-left panel, gives the
2D likelihood distribution of metallicity against light-weighted age
(adjusted to $z=0$), with the degeneracy manifesting itself as an
apparent anti-correlation between the two parameters. The bulk of the
distribution is centred on ages of 6--7~Gyr and [Z/H]$\sim$0.2,
reflecting the 1D distributions in Figure~\ref{fig:agezmass}.  To
quantify this degeneracy we calculate the median metallicity in each
of the age bins of Figure~\ref{fig:twofinal} and fit a straight line
through the resulting points. This gives a slope of $\Delta \log
Z/\Delta \log t = -1.05$~dex yr$^{-1}$.

Previous studies \citep{caldwell-2003-125,kelson-2006-642} have
suggested possible correlations between the age and metallicity of
early-type galaxies with velocity dispersion,
$\sigma$. \citet{kelson-2006-642} go so far as to claim that the
properties of early-type galaxies in clusters can be completely
characterised as a one-parameter family, given by the velocity
dispersion of the galaxy. The Faber-Jackson relation also motivates
investigation into the dependence of mass on velocity dispersion. The
right-hand column of Figure~\ref{fig:twofinal} shows how the three
parameters, age, metallicity and mass, depend on velocity dispersion
in our sample. The age seems to show a small correlation with velocity
dispersion:\ the median yields a slope of $\Delta$age$/\Delta \sigma
\sim 4$~Myr/kms$^{-1}$. The metallicity shows an anti-correlation with
$\Delta [Z/H]/\Delta \sigma \sim - 1 \times 10^{-3}$~dex/kms$^{-1}$.
The mass shows a slight positive correlation of $\Delta
\log(M/M_\odot)/\Delta \sigma \sim 9 \times
10^{-4}$~dex/kms$^{-1}$. The modes of the distribution, however, are
well concentrated, centred at about $240\,{\rm km\, s^{-1}}$.

\subsection{Star formation history of LRGs}
\label{subsec:sfh}

In addition to deriving the rest-frame parameters of these galaxies,
we would like to constrain their star formation histories by tracking
the overall star formation rate as a function of cosmic time.  To
achieve this, we compute the lookback time for each galaxy and
extrapolate the past and future star formation histories for each
model, weighted in the usual way. This gives us a PDF for star
formation rate, in bins of lookback time, which can be summed over the
entire sample to give an overall likelihood distribution. It should be
noted that this is only done for the continuous component of the
models, and does not include star formation from the burst.  We also
note that the earlier studies of \kau{} and \gal{} used Monte Carlo
libraries generated with $0 < \gamma < 1$ for the continuous
component, while here we choose $0 < \gamma < 2$. We discuss below the
impact of changing this prior on the derived star formation history.

\begin{figure}
\includegraphics[angle=-90,width=8cm]{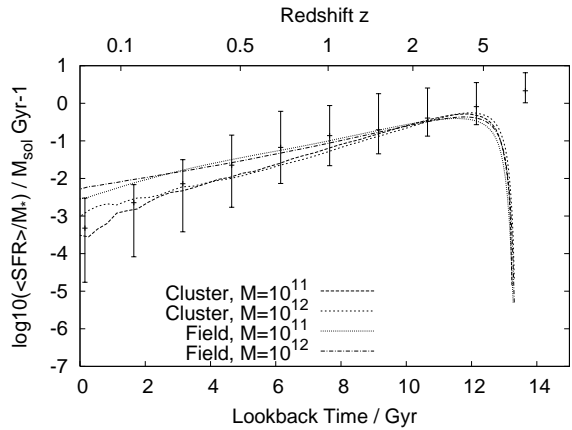}
\caption{The median star formation rate of LRGs in our sample, binned
by lookback time. (See text for details.) The continuous lines are
the semi-analytic model predictions of \citet{delucia-2006-366}.
\label{fig:cosmicsfr}}
\end{figure}

In Figure~\ref{fig:cosmicsfr}, we present the median values of star
formation rate per stellar mass, binned by lookback time. The error
bars show the 16th and 84th percentile limits of the distributions in
each bin. Also shown is the formation rate from the
\citet{delucia-2006-366} models (from their figure~1) with stellar
masses of $10^{11}\,M_\odot$ and $10^{12}\,M_\odot$, for galaxies in
clusters and in the field. For reference, the LRG sample used here is
selected in the redshift range $0.15 < z < 0.4$, corresponding to
lookback times of $1.8~\textrm{Gyr} < t_{lb} < 4.2~\textrm{Gyr}$.

Between lookback times of 3 and 12~Gyr, the LRG results and the
simulation show good agreement, with all points lying within the
68-percentile range of the predicted value. For shorter lookback
times, the data give a lower star formation rate than expected when
compared with simulated field galaxies. In this regime of lookback
time, we are extrapolating the future star formation of a galaxy using
the passive exponential model and take no account of any possible
interactions. This may account for the discrepancy, as additional
mergers may boost the star formation rate. Interestingly, there is
much better agreement at more recent times when considering galaxies
in clusters. This is consistent with the inference that most LRGs
reside in clusters \citep{1989ARA&A..27..235K}. At lookback times
exceeding 12~Gyr, the model curves drop off sharply, while the data
still reveal some residual star formation. This is because we only
show star formation rates for galaxies with stellar populations
existing at these early times. Accordingly, the measured star
formation rate at 12~Gyr accounts for only a small fraction of
galaxies that will be created in the Universe.

When considering the overall formation mechanism of elliptical
galaxies, it is important to distinguish between the \emph{formation
times} of the composite stars and the \emph{assembly time} of the
constituent progenitors. The formation time of a system can be defined
as the time when a certain fraction of stellar mass has been formed,
compared with the mass at $z=0$. Here we calculate the cosmic lookback
times corresponding to the redshift at which star formation first
begins, and when 50 per cent and 80 per cent of the stellar mass has
been formed, denoted Tform, Tf50 and Tf80 respectively. These
parameters give us not only a handle on when the stars in the galaxies
formed, but the time-scale over which the stars which reside in LRGs
today were forming.

\begin{figure}
\includegraphics[width=8cm]{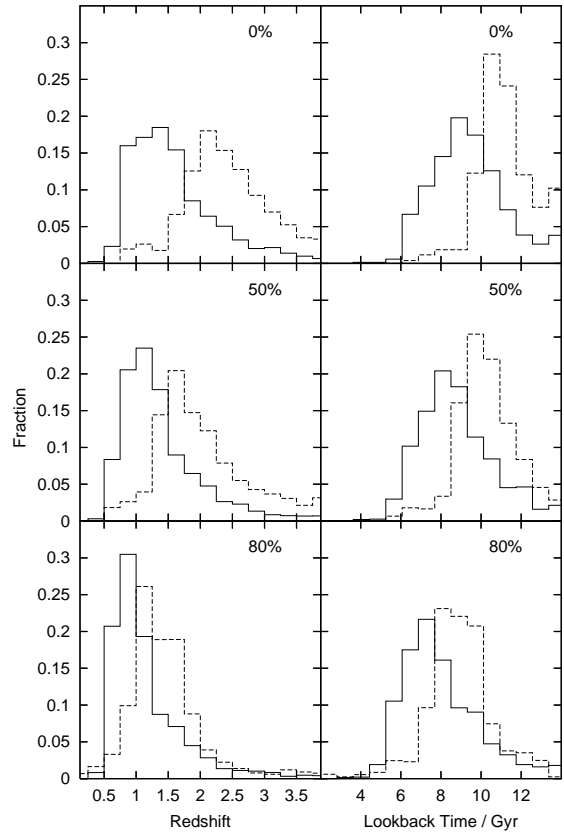}
\caption{The redshift (left-hand column) and lookback time (right-hand
column) at the formation time and when 50 per cent and 80 per
cent of the stellar mass is in place, compared with the mass at
$z=0$. (Top, middle and bottom panels respectively.) The solid line
indicates that distribution for a library in which $0.0 < \gamma <
2.0$, while the dotted line corresponds to $0.0 < \gamma < 1.0$. The
lower two distributions (50 per cent and 80 per cent) are comparable
to Figure~4 of \citet{delucia-2006-366}.
}
\label{fig:tform}
\end{figure}

The solid histograms in Figure~\ref{fig:tform} show the distributions
of these formation times and their corresponding
redshifts. Table~\ref{tab:tform} summarizes these results, giving the
lower quartile, median and upper quartile values. The results show
that the stellar component of LRGs typically first began forming
around 8-10~Gyr ago ($z \approx 1.1-1.9$), with 80 per cent of the
stellar mass in place 6-8~Gyr ago ($z \approx 0.7-1.1$).

These values are directly comparable with those provided in table~1 of
\citet{delucia-2006-366}, whose semi-analytic models predict
$\textrm{Tf50} \sim 11$~Gyr and $\textrm{Tf80} \sim 9-10$~Gyr. Our
results for $0<\gamma<2$ give ages about 2--3~Gyr younger than the
simulation predictions.

\begin{table}
\begin{tabular}{|c|r|r|r|r|}
\hline
&& 25\% & 50\% & 75\% \\
\hline
\multirow{3}{*}{$0 < \gamma < 2$} & Tform & 8.15 & 9.27 & 10.59 \\
& Tf50 & 7.52 & 8.58 & 9.97 \\
& Tf80 & 6.63  & 7.63  & 9.12 \\
\hline
\multirow{3}{*}{$0 < \gamma < 1$} & Tform & 10.38 & 11.08 & 12.14 \\
& Tf50 & 9.37 & 10.17 & 11.22 \\
& Tf80 & 8.03  & 8.90  & 9.74 \\
\hline
\citet{delucia-2006-366} & Tf50 & 9.78 & 10.56 & 11.22 \\
($10.8 < \log(M/M_{\sun}) < 11.5$)& Tf80 & 7.56  & 9.20  & 10.31 \\ \hline
\citet{delucia-2006-366} & Tf50 & 11.01 & 11.41 & 11.76 \\
($11.5 < \log(M/M_{\sun}) < 12.2$) & Tf80 & 9.50  & 10.31  & 10.79 \\
\hline
\end{tabular}
\caption{The lower quartile (25 per cent), median (50 per cent) and
upper quartile (75 per cent) values summarizing the lookback times (in Gyr)
since formation of LRGs in our sample. Values are shown for the full
Monte Carlo library, for which $0 < \gamma < 2$, and also the results
when the library is restricted to models with $0 < \gamma < 1$. For
reference, the corresponding values from table~1 of
\citet{delucia-2006-366} are reproduced here for comparison.}
\label{tab:tform}
\end{table}

\begin{figure}
\begin{tabular}{cc}
   \includegraphics[angle=-90,width=8cm]{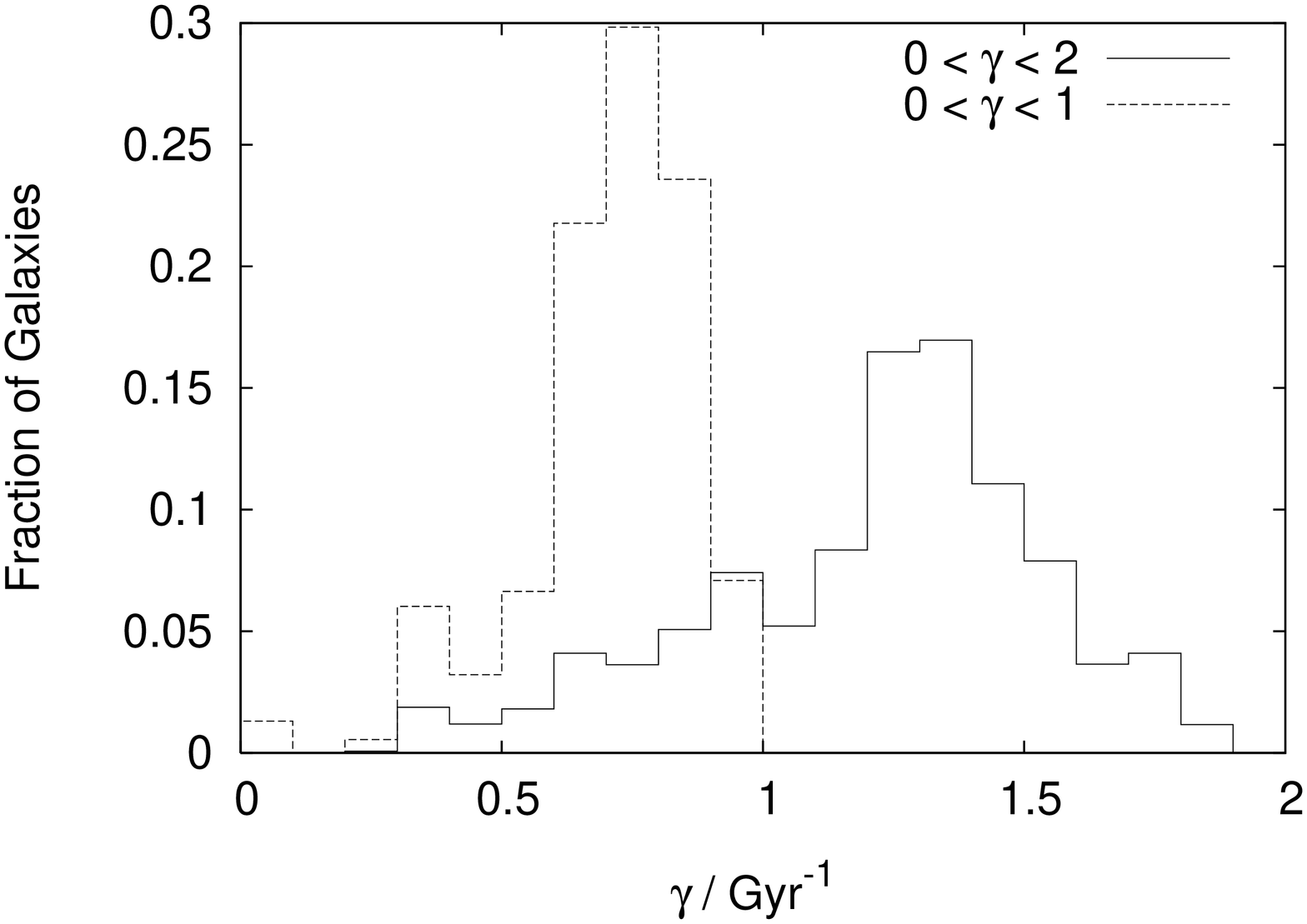} \\
   \includegraphics[angle=-90,width=8cm]{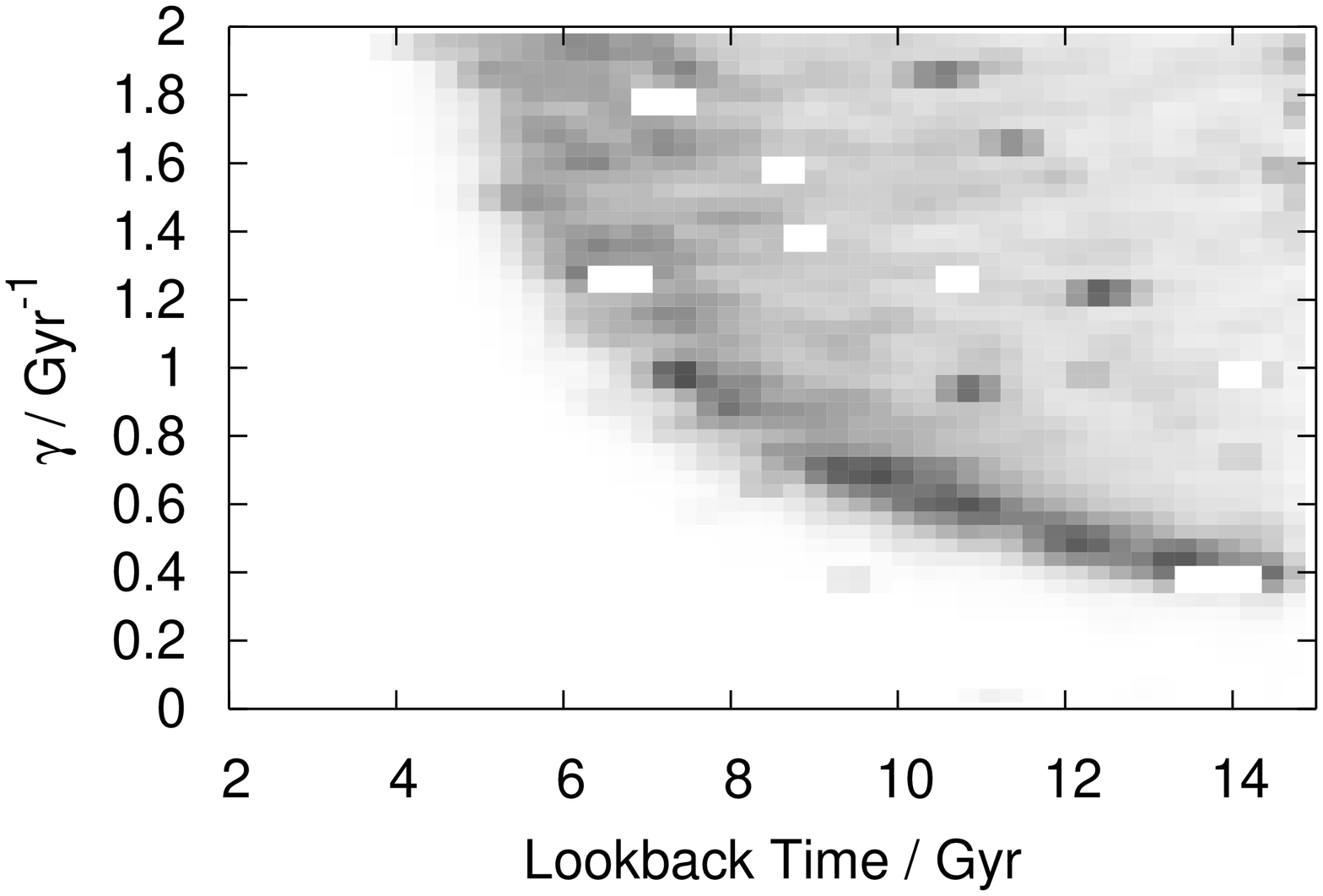}
\end{tabular}
\caption{The top panel shows the distribution of $\gamma$ for the LRG
sample when the prior range is $0.0 < \gamma < 2.0$ (solid) and $0.0 <
\gamma < 1.0$ (dashed). The lower panel shows the 2D PDF for $\gamma$
as a function of the lookback time corresponding to the time of formation.}
\label{fig:gamma}
\end{figure}

In Figure~\ref{fig:gamma} (top), we compare the distribution of the
median $\gamma$-values when $0 < \gamma < 2$ is allowed and when the
library is restricted to models for which $0 < \gamma < 1$. The
difference in predicted values is striking, with median values of 1.28
and 0.73 for the two distributions, respectively. The impact that this
has on the predicted star formation history is shown in
Figure~\ref{fig:gamma} (bottom). This plot gives the 2D probability
distribution of $\gamma$ as a function of the time since formation of
the galaxy, produced in the same way as Figure~\ref{fig:twofinal}. We
see here that the two parameters appear to be inversely proportional
to one another, with older objects having the most prolonged periods
of star formation. Generally, only objects older than 7~Gyr are
expected to have $\gamma < 1$. The overall effect of restricting the
prior on $\gamma$ to $0< \gamma < 1$ is therefore to increase the
derived ages of the galaxies.

This is illustrated by the dashed histograms of Figure~\ref{fig:tform}
and the corresponding values in Table~\ref{tab:tform}. Here we see
that for the more restricted range in $\gamma$, the predicted median
formation time is increased by 1.8, 1.6 and 1.3~Gyr for Tform, Tf50
and Tf80 respectively. The inter-quartile range for the formation time
distributions are reduced from 2.4~Gyr ($0 <\gamma < 2$) to 1.7~Gyr
($0 <\gamma < 1$). This is understandable as by reducing the range of
$\gamma$, we are also reducing the number of possible star formation
histories for a given formation time. Using the more restricted range
of $\gamma$ brings our predictions in closer agreement with those of
\citet{delucia-2006-366}. We note that restricting the prior value of
$\gamma$ also increases the median light-weighted age by 1-2~Gyr, but
appears to have little effect on the derived metallicity and mass
distributions.

\begin{figure}
\includegraphics[width=8cm]{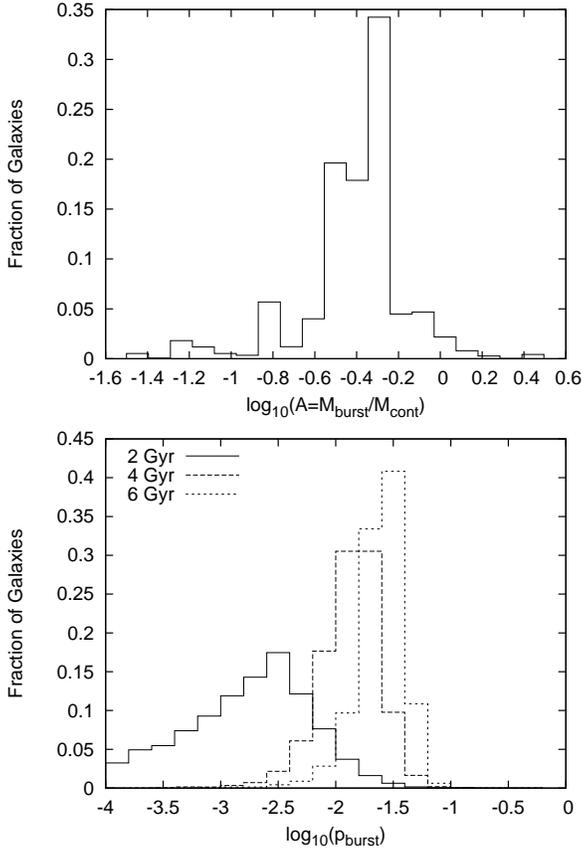}
\caption{The top panel shows the distribution of $A$, the ratio of
mass formed in a burst compared with that formed by the exponentially
decaying model. The lower panel shows the distribution of
burst probabilities. The solid, dashed and dotted lines show the
distribution of $p_{\rm burst}$, the probability that a galaxy
experienced a burst, in the last 2, 4 and 6~Gyr, respectively.}
\label{fig:burst}
\end{figure}

In addition to tracking the formation times of galaxies, it is
important to discover whether galaxies have undergone any recent
bursts of star formation activity. For each galaxy, we calculate the
probability $p_{\rm burst}$ of a burst by adding the weights of every
model that has experienced a burst in the last $n$~Gyr, and
normalising by the total weight of all models in the library. The
distribution of burst probabilities for $n=2,4,6$ is shown in
Figure~\ref{fig:burst} (bottom panel). The probabilities are generally
small. Typically there is only a 0.25 per cent chance of a burst
occuring within the last 2~Gyr for an individual galaxy. This rises to
1.6 per cent for bursts within the last 4~Gyr and 3.2 per cent within
6~Gyr. We note that the burst probabilities may be sensitive to the
prior. For reference, we set up the library so that 1 per cent of
galaxies experience a burst within the last 2~Gyr. We would also like
to calculate the magnitude of these bursts compared with the
underlying continuous model. To do this we consider only those models
in which bursts have occured within the lifetime of the galaxy and
calculate the median value of $A=M_{\rm burst}/M_{\rm cont}$ by
marginalising in the usual way. The top panel of
Figure~\ref{fig:burst} gives a typical value of $A=0.4$, implying that
around 29 per cent of the total mass of stars in a galaxy that had a
burst in its history would have come from the burst.

\subsection{Metallicity of LRGs}\label{subsec:z}

In this section, we discuss implications of the measured metallicity
distribution of the LRGs in terms of their chemical evolution.
The simplest possible model for the chemical evolution of a galaxy is the
closed-box model of \citet{talbot-1971-170}. We also consider a standard
modification allowing for continued accretion of gas by the galaxy.
Details of both models are discussed in \citet{BM98}. 

In the closed box model, it is assumed that initially the material is
entirely gaseous, possibly enriched to a metallicity level $Z_i$. As
the name suggests, no material is allowed to enter or leave the
region, so there is no further appreciable interaction of the galaxy
with its surroundings once it has formed. As the galaxy evolves, stars
are formed, consuming interstellar gas in the process and polluting
the remaining gas with heavy elements corresponding to a yield $p$ of
mass in metals per mass of stars formed. The yield is assumed
constant. It is further assumed that turbulence keeps the gas
homogeneous so that the instantaneous recycling approximation may be
made, which neglects the delay between the formation of a population
of stars and the ejection of heavy elements into the remaining gas by
massive stars in the population.

\begin{figure}
\begin{tabular}{cc}
   \includegraphics[angle=-90,width=8cm]{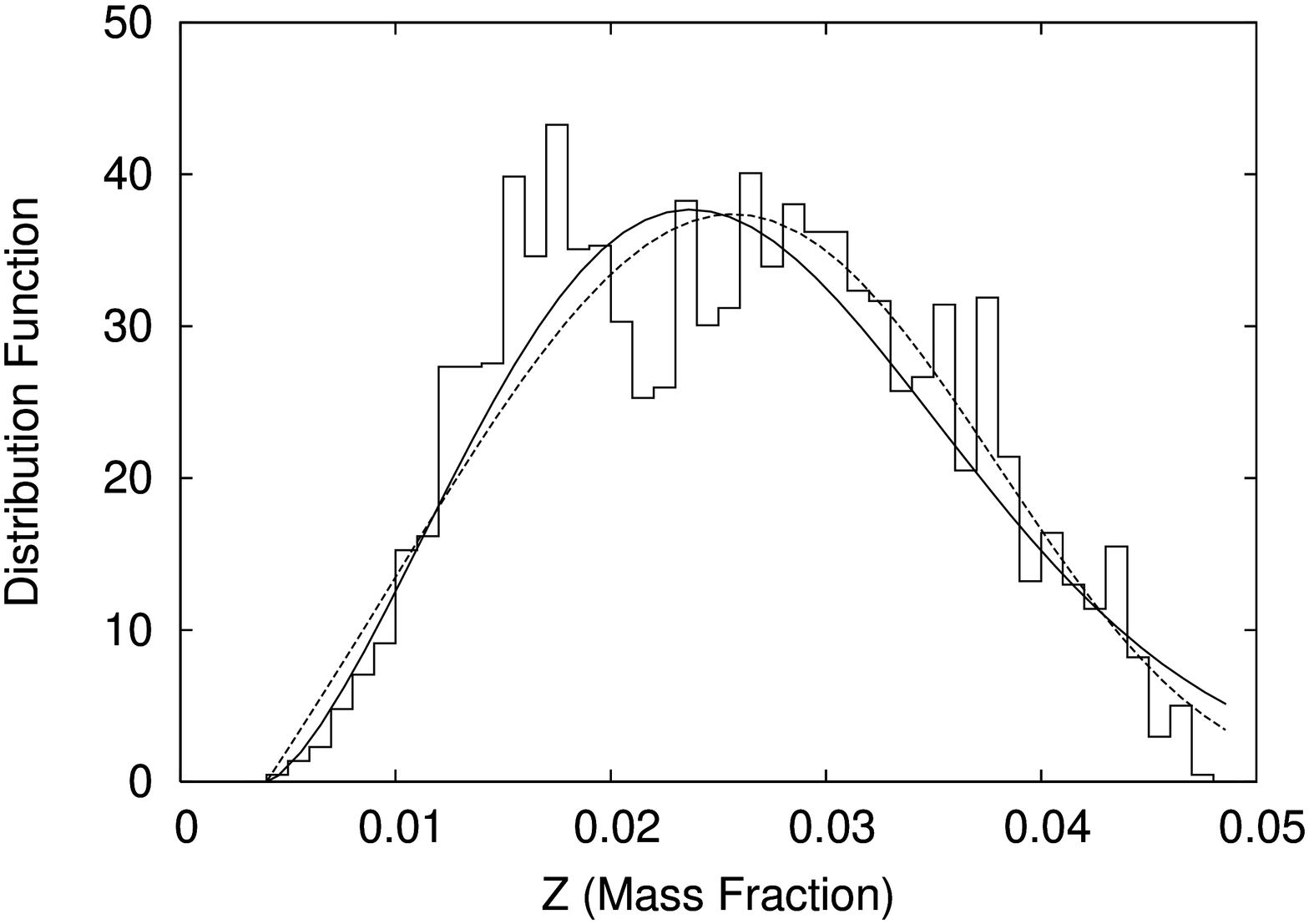} \\
   \includegraphics[angle=-90,width=8cm]{fig11b.eps}
\end{tabular}
\caption{The top panel shows the metallicity distribution function for
LRGs in our sample (histogram). The solid line is a fit of a closed
box SSQ model, while the dashed line is for an accreting box SSQ model
(see text). The bottom panel shows the relation between age and
metallicity predicted by the accreting box SSQ model for
characteristic quenching time-scales of 0.65~Gyr (solid lines) and
1.3~Gyr (dashed lines) for formation times of 9~Gyr and 12~Gyr.}
\label{fig:zdis}
\end{figure}

Suppose the interstellar gas has a mass ${\cal M}_g$ and contains a mass
of heavy elements ${\cal M}_h$. The metallicity of the gas is then
\begin{equation}
Z = \frac{{\cal M}_h}{{\cal M}_g}.
\end{equation}
In this model, the metallicity may be shown to evolve according to
\begin{equation}
Z(t) = Z_i - p\ln \left[\frac{{\cal M}_g(t)}{{\cal M}_g(0)}\right].
\label{eq:zev_cb}
\end{equation}
The mass of stars in the box with metallicity between $Z_i$ and $Z(t)$ is then
\begin{eqnarray}
{\cal M}_s(<Z) &=& {\cal M}_s(t) = {\cal M}_g(0) - {\cal M}_g(t) \nonumber \\
               &=& {\cal M}_g(0)\left[1-\exp\left(-\frac{Z-Z_i}{p}\right)\right],
\label{eq:box}
\end{eqnarray}
and the metallicity distribution of the stars is
\begin{equation}
\frac{d{\cal M}_s(<Z)}{dZ}=\frac{{\cal M}_g(0)}{p}
\exp\left(-\frac{Z-Z_i}{p}\right)
\label{eq:dMSdZ_box},
\end{equation}
for $Z>Z_i$, and vanishes for $Z<Z_i$.
The mean metallicity of the stars is then
\begin{eqnarray}
\bar Z(t) &=& \frac{1}{{\cal M}_s}\int_{Z_i}^{Z(t)}\,Z
\frac{d{\cal M}_s}{dZ}\,dZ \nonumber \\
          &=& Z_i + p\left(1+\frac{x\ln x}{1-x}\right),
\label{eq:Zbar}
\end{eqnarray}
where $x=1-{\cal M}_s(t)/{\cal M}_g(0)=\exp[-(Z-Z_i)/p]$. The limiting
metallicity values of $Z=Z_i$ and $Z\rightarrow\infty$ correspond
respectively to $x=1$ and $x=0$, and to $\bar Z=Z_i$ and $\bar
Z=Z_i+p$. The mean metallicity is thus restricted to being smaller
than the sum of the initial metallicity and the yield. Inspection of
Figure~\ref{fig:zdis} suggests the yield must then be as high as
$p\approx0.05$.

In this scenario, the distribution of mean metallicity is a reflection
of the distribution of initial metallicities $Z_i$ of the galaxies and
their ages, with younger galaxies expected to have lower metallicities
on average. However, this picture is not supported by the
age-metallicity plot of Figure~\ref{fig:twofinal}, which, while
showing a large amount of scatter for super-solar metallicity values,
also shows a general anti-correlation, particularly at low metallicity
values. This effect is likely in part due to a residual
age-metallicity degeneracy, although the degree to which the
degeneracy masks a true underlying trend is unclear.

We next consider a modification of the closed box model in which star
formation is abruptly halted after an enrichment phase. The model is
motivated by the proposed solution to the anti-hierarchical behaviour
of early type galaxies invoking a sudden quenching of star formation
in the sub-units which eventually coalesce into the galaxies we see
today. In this model, we allow for {\it stochastic star-formation
quenching} (SSQ), modelled as a Poisson process at a uniform rate
$\lambda$. The probability that star formation is suddenly quenched in
any given galaxy between the times $t$ and $t+dt$ is then $\lambda
dt$. The metallicity level any galaxy achieves is then given by the
time during which the gas was enriched before star formation ceased,
as well as by any initial metallicity. The metallicity distribution is
given by
\begin{equation}
\frac{1}{{\cal N}_{\rm tot}}\frac{d{\cal N}}{d\bar Z}=e^{-\lambda t}
\lambda\frac{dt}{d\bar Z}.
\label{eq:dNdZ}
\end{equation}
The time derivative $d\bar Z/dt$ requires specifying a star formation
rate $\psi(t)$. The simplest approximation is a constant rate $\psi(t)=\psi_0$.
For reasons explained below, we instead consider the more general form
\begin{equation}
\psi(t) = \psi_\beta t^{-\beta},
\label{eq:psi}
\end{equation}
where $\psi_\beta$ is a constant and $0\le\beta<1$.  A constant rate
corresponds to the special case $\beta=0$. For $\beta>0$, the rate
corresponds to a burst at $t=0$, gradually decaying afterwards. The
restriction $\beta<1$ ensures a finite mass of stars is formed.

Expressing $dt/d\bar Z=(dt/dx)/(d\bar Z/dx)$ and noting
$\psi(t)=d{\cal M}_s/dt$ gives
\begin{equation}
x(t) = 1 - \frac{\psi_\beta}{{\cal M}_g(0)(1-\beta)}t^{1-\beta},
\label{eq:xev}
\end{equation}
and
\begin{equation}
\frac{p}{{\cal N}_{\rm tot}}\frac{d{\cal N}}{d\bar Z}=-\frac{\alpha}{1-\beta}
\exp\left[-\alpha(1-x)^{1/(1-\beta)}\right]\frac{(1-x)^\frac{2-\beta}{1-\beta}}
{1-x+\ln x},
\label{eq:pdNdZ_cb}
\end{equation}
where $\alpha=\lambda[{\cal
M}_g(0)(1-\beta)/\psi_\beta]^{1/(1-\beta)}$ is the ratio of the gas
consumption time-scale to quenching time-scale. Typically this ratio is
expected to be large, otherwise most of the mass will have been
enriched to its maximum mean value through generations of star
formation before star formation ceases.

In the limit of small mean metallicity values, $\bar Z\rightarrow
Z_i$, the distribution function $(p/{\cal N}_{tot})d{\cal N}/d{\bar
Z}\rightarrow 2\frac{\alpha}{1-\beta}(1-x)^{\beta/(1-\beta)}$ for
$x\rightarrow1$. For the special case $\beta=0$, the limiting value is
$2\alpha$. This limiting value will in fact apply to any analytic form
for $\psi(t)$ that is non-vanishing at $t=0$, including an
exponentially decaying star formation rate. Such asymptotic behaviour,
however, is contrary to the diminishing distribution at low
metallicity values in Figure~\ref{fig:zdis}. For $0<\beta<1$, the
distribution function vanishes for $\bar Z\rightarrow Z_i$. This is
the reason for having introduced this particular `cuspy' form for the decaying
burst.

We next modify the simple closed box model by assuming that metal free
gas may be accreted by the galaxy from its surroundings. The change in
total mass is $\delta{\cal M}_t = \delta{\cal M}_s + \delta{\cal M}_g
\ne 0$. In this case, the metallicity evolves according to
\begin{equation}
Z(t) = p\left[1-\left(1-\frac{Z_i}{p}\right)e^{x-1}\right],
\label{eq:zev_ab}
\end{equation}
where $x=1-M_s(t)/ M_g(0)$, as for the closed box case. The relation
between $x$ and $t$ is again given by Eq.~\ref{eq:xev}. As there is no
upper limit to the amount of gas accreted, there is no upper limit to
the amount of stars which may form. As a consequence, $x$ may become
arbitrarily small. In the limit $x\rightarrow -\infty$, the
metallicity takes on the value of the yield ($Z\rightarrow p$).  The
maximum metallicity achievable is the yield, provided $Z_i<p$.  (For
$Z_i>p$, the metallicity will decrease with time towards $p$.)

The mean metallicity evolves according to
\begin{equation}
{\bar Z}(t) = p + \frac{Z(t) - Z_i}{\ln\frac{p-Z(t)}{p-Z_i}},
\label{eq:Zavg_ab}
\end{equation}
with the limiting value $\bar Z\rightarrow p$ as $x\rightarrow
-\infty$. In analogy to the closed box case, the metallicity
distribution in the SSQ model is given by
\begin{eqnarray}
\frac{p}{{\cal N}_{\rm tot}}\frac{d{\cal N}}{d\bar Z}&=&\frac{\alpha}{1-\beta}
\exp\left[-\alpha(1-x)^{1/(1-\beta)}+1-x\right] \nonumber \\
&&\times\frac{(1-x)^\frac{1}{1-\beta}}
{\frac{Z-Z_i}{1-x}e^{1-x} + Z_i - p}.
\label{eq:pdNdZ_ab}
\end{eqnarray}

Best fits to the measured distribution are shown in
Figure~\ref{fig:zdis} (upper panel) for both the closed box (solid
line) and accreting box (dashed line) SSQ models. For the closed box
SSQ model, the parameters for the fit are $Z_i=0.0040$, $\alpha=66$,
$\beta=0.59$ and $p=0.26$. For the accreting box SSQ model, the
parameters for the fit are $Z_i=0.0040$, $\alpha=2.3$, $\beta=0.50$
and $p=0.10$. Both models well describe the distributions. Error bars
are not quoted because neither fit has an acceptable value of $\chi^2$
(reduced $\chi^2\approx 5$), due in large part to non-Poisson
fluctuations from bin to bin. A dip near solar metallicity hints at a
bimodal distribution. Without extensive Monte Carlo tests of the
population synthesis model-fitting procedure, it is difficult to
assess the statistical significance of the fluctuations, and of any
evidence for bimodality. Doing so is beyond the scope of this paper,
but may be worth considering in any future estimates of the galaxy
metallicity distribution. The spread in parameter values between the
two models provides some indication of the errors. We note that the
excess number density of systems predicted by the models for
$Z\approx0.05$ may be an artefact of the stellar library, which is
restricted to $Z\le0.05$.

Values for $\alpha$ exceeding unity are expected, and confirm the
assumption that star formation is quenched on a time-scale short
compared with the time-scale for complete conversion of the initial
gas mass into stars. Adopting two e-foldings of star formation in the
population synthesis models for the conversion time, the distribution
of $\gamma$ in Figure~\ref{fig:gamma} suggests a typical time-scale
for depleting the initial gas content of $2/\gamma\sim
1.5\times10^9$~yr. The corresponding quenching time-scale
$\lambda^{-1}$ for the closed box SSQ model is then
$2/\gamma\alpha\approx2\times10^7$~yr, comparable to the Salpeter time
($t_{\rm Salp}=\epsilon\sigma_{\rm T}c/ 4\pi Gm_{\rm p}$) for a black
hole energy conversion efficiency $\epsilon$ of 5 per cent, and to the
lower limit for the lifetime of a luminous QSO
\citep{2002MNRAS.335..965Y}. The effective yield value for the model
is quite high, comparable to that expected from massive stars alone
\citep{1992A&A...264..105M,2005A&A...433.1013H}, and so is not
expected for a normal initial mass function. Such an IMF would result
almost entirely in stars sufficiently massive to produce Type II
supernovae. A metallicity value corresponding to Type II supernovae
ejecta is consistent with the evidence for $\alpha$-enhanced
abundances. An IMF dominated by massive stars should not necessarily
be precluded. An initial mass function heavily weighted towards
massive stars ($dN/dm\sim 1/m$) has been invoked in recent
semi-analytic galaxy formation models to reproduce sub-millimetre
galaxy counts and the high metallicities observed in the intracluster
gas of rich galaxy clusters (Lacey et al. 2006, in
preparation). Luminous Red Galaxies are believed to reside
predominantly in the cores of just such rich clusters.

For the accreting box SSQ model, the smaller value for $\alpha$
corresponds to a star-formation quenching time-scale of
$\sim6\times10^8$~yr, comparable to the upper limit for the lifetime
of a luminous QSO \citep{2002MNRAS.335..965Y}. The effective yield
value for the model is still high, about twice that expected for the
Chabrier (2003) initial mass function for early stars, for which a
little under half of the stellar mass is contained in stars massive
enough to form Type II supernovae, and assuming an approximate mean
ejected mass fraction of 10 per cent. It suggests again an IMF tilted
toward more massive stars, but not as strongly as required by the
closed box model.

The sudden star-formation quenching scenario may also account in part
for the inverse correlation between age and metallicity. For a given
formation time, systems with an extended amount of star formation
before quenching will produce stars with increasing metallicity and a
younger average population compared with systems in which star
formation was quenched early. Because of the short characteristic
decay time for star formation, however, the ages of the stars will not
differ greatly (typically $\sim2\gamma^{-1} \approx 1.5$~Gyr), so that
the inverse relation must also result from a range in formation
times. For slower star formation, however, a broader range in ages is
possible. The relation between the lookback time to the time at which
star formation was quenched and the mean metallicity of the stars at
that time is shown in Figure~\ref{fig:zdis} (lower panel) for the
accreting box SSQ model. Values of $\lambda^{-1}=0.65$~Gyr (solid
lines) and $\lambda^{-1}=1.3$~Gyr (dashed lines) are shown, with
formation times of 9~Gyr and 12~Gyr. These curves span the range of
metallicity vs age in Figure~\ref{fig:twofinal}. The rapid ascent at
low (sub-solar) metallicities, and slower metallicity evolution at
higher (super-solar) metallicities, predicted by the model suggests
that most galaxies will be concentrated in the region of super-solar
metallicities, consistent with the 2D age-metallicity distribution in
Figure~\ref{fig:twofinal}.

\section{Discussion \& Conclusions}
\label{sec:discussion}

In this section, we compare our results with those in previous works,
and summarise the main conclusions of our study. The groups at
MPA/JHU have made publically available large catalogues of the results
of model fits to SDSS data. These include the mass catalogues of
\kau{} (updated to Data Release 4) and age, mass and metallicity
values of \gal{} (Data Release 2)\footnote{Data available
electronically at {http://www.mpa-garching.mpg.de/SDSS}}.

\begin{figure*}
\includegraphics[width=15cm]{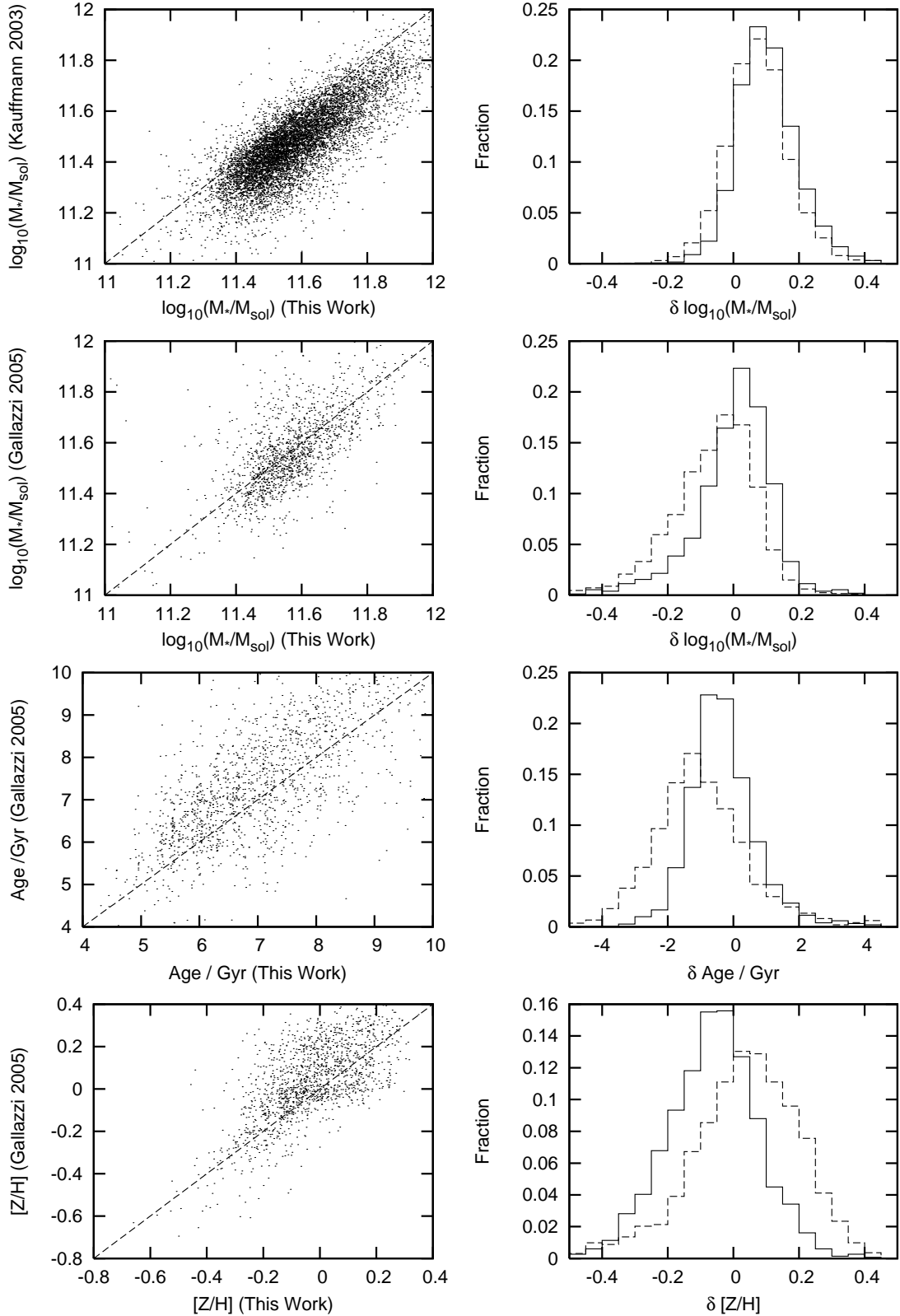}
\caption{Comparison of results from this paper with previous work. The
left-hand column shows our stellar mass values against those of K03
and G05 (top two plots), and our age and metallicity against the
values of G05 (bottom two plots). The values quoted have been
calculated by fitting the similar indices and priors as K03 and G05
respectively. The right hand column shows the residual between our
values and the corresponding previous result. The solid histograms use
the same indices and similar priors as K03 or G05, while the dashed
histograms use our set of indices and Monte-Carlo priors.}
\label{fig:comp}
\end{figure*}

To make our results comparable with values from \kau{} and \gal{}, the
library is restricted to models with $0 < \gamma < 1$, to more closely
match our prior with theirs. We then re-compute masses by fitting only
the D$_n(4000)$ and H$\delta_A$ indices to allow direct comparison
with K03. Similarly, we calculate masses, ages and metallicities by
fitting the $[\textrm{Mg}\textrm{Fe}]'$, $[\textrm{Mg}_2\textrm{Fe}]$,
H$\beta$, H$\gamma_A$+H$\delta_A$ and D$_n(4000)$ indices, as used in
G05. We also match cosmologies, adopting a flat
Friedmann-Robertson-Walker universe with $\Omega_m =0.3$ and $H_0 =
70~\textrm{kms}^{-1}\textrm{Mpc}^{-1}$.

The left-hand column of Figure~\ref{fig:comp} shows a comparison
between our values and those from the MPA/JHU catalogue. All derived
parameter values show generally good consistency over the ranges
studied. The right hand-column of Figure~\ref{fig:comp} shows the
distribution of differences between our values and the MPA/JHU
values. The solid histogram uses the modified prior and K03 or G05
indices as above. The dashed histogram shows the same residuals using
the full library ($0 < \gamma < 2$) and the original set of indices
($[\textrm{Mg}\textrm{Fe}]'$,$[\textrm{Mg}_1\textrm{Fe}]'$, H$\beta$,
H$\gamma_F$ and D$_n(4000)$).

The stellar mass distribution we derive is narrowly peaked around
$3\times10^{11}\,M_\odot$. A comparison between our derived stellar
masses and those of K03 shows overall consistency, with a scatter of
$\sim 0.16$~dex. Our stellar mass values are in general larger than
theirs by $\sim 0.09$~dex when using their indices, and $\sim
0.07$~dex using ours. There is a smaller offset when we compare our
masses to those derived by G05:\ $\sim 0.02$~dex using their set of
indices, and $\sim -0.04$~dex with ours.

The offsets are likely to be due to a combination of
factors. Principal among these is the difference in the adopted
initial mass function (IMF). While we use the Salpeter IMF, K03 use
the \citet{2001MNRAS.322..231K} IMF and G05 use that of
\citet{2003PASP..115..763C} for disc stars. \citet{bruzual-2003-344}
show that, for a simple stellar population with solar metallicity, the
Salpeter IMF results in an increased stellar mass-to-light ratio of
about 0.2~dex over the Kroupa and Chabrier IMFs, at a given
evolutionary time. Our stellar mass values agree with the results of
K03 and G05 more closely than this, which on the surface is
surprising. The reason for the good agreement appears in fact to be in
part fortuitous. The effect of broadening the range of $\gamma$ to $0
< \gamma < 2$ is to find ages that are, on average, $\sim1.1$~Gyr
smaller than those of K03 and G05. If we limit the prior to $0 <
\gamma < 1$, this difference reduces to $\sim0.4$~Gyr. The scatter in
ages is around $\sim 1$~Gyr. According to the results of
\citet{bruzual-2003-344}, the larger ages found by K03 and G05
correspond to an increase in the mass-to-light ratio of close to
0.1~dex over the values corresponding to our best fits. This
difference brings our stellar mass estimates and those of K03 and G05
into fairly close agreement. Had we adopted the
\citet{2001MNRAS.322..231K} or \citet{2003PASP..115..763C} IMF, the
peak in the measured stellar mass distribution would be smaller by
about 0.2~dex, or $2\times10^{11}\,M_\odot$.

When using $0 < \gamma < 2$, we also find that our best fitting models
are more metal-rich compared with those of G05 by $\sim 0.05$~dex,
with a scatter of $\sim 0.15$~dex. Our models, however, are
comparatively metal-poor by $\sim 0.05$~dex when using $0 < \gamma <
1$. Here we see the age-metallicity degeneracy in effect:\ younger
ages correlate with higher metallicities, and vice versa.

Sources of discrepancy in addition to the difference in priors include
the rejection by K03 of models for which the $r-i$ colour is redder
than observed, the difference in the population synthesis techniques
adopted to generate the models and the consequent difference in the
abilities of the models to reproduce the absorption line features in
the spectra. We include models with $A_z$ attenuations down to $-0.1$,
following the method of G05. We also note that we are using a
different extinction curve to that adopted by K03 and G05, who both
use the single power law ($\propto \lambda^{-0.7}$) of
\citet{calzetti-1994-429}. The derived $A_z$ values we find are
sufficiently small (see Section~\ref{subsec:pe}) to have a negligible
effect on the mass estimates.

One way to examine the effect of these different choices is to examine
how well the two models reproduce the absorption indices. In figure~18
of \citet{bruzual-2003-344}, the H$\beta$, H$\delta_A$, H$\gamma_A$,
D$_n$(4000), ${\rm [MgFe]'}$, [Mg$_1$Fe] and [Mg$_2$Fe] index
strengths are used to fit high signal-to-noise galaxies from the SDSS
Early Data Release (EDR). A comparison of this plot with our
Figure~\ref{fig:lick} shows general consistency in the systematic
offsets of many indices, despite the broader range in galaxy types
modelled by \citet{bruzual-2003-344}. For example, both models fit the
H$\beta$, D$_n$(4000) and ${\rm [MgFe]'}$ well, whilst underestimating
the strengths of the CN$_1$, CN$_2$, Ca4455, C$_2$4668, Mg$_b$ and NaD
indices and overestimating Fe4383, Fe5015, Fe5270 and Fe5335. This
behaviour is encouraging as it allows us to check the consistency of
both models. There are some indices that are not consistent, however.
For example, we tend to overestimate the strengths of the two
magnesium indices, Mg$_1$ and Mg$_2$, while the
\citet{bruzual-2003-344} models underestimate them. Other indices that
show differences are Ca4227, G4300, Fe4531, Fe5406, TiO$_1$, TiO$_2$
and H$\delta_A$. These differences are likely in part due to the
differences in the galaxy populations modelled.

The results of Section~\ref{subsec:sfh} highlight an important
consideration when setting up this type of analysis, namely what range
should $\gamma$ take in the prior? We have shown here that doubling
the upper limit of $\gamma$ from 1 to 2 decreases the formation
redshift considerably from $z\sim2.0$ to $z\sim1.0$, making the
derived ages younger by around 2~Gyr. This anti-correlation between
$\gamma$ and time since formation shown in Figure~\ref{fig:gamma} is
an example of the degeneracy between formation epoch and formation
time-scale. So an old, slowly-forming galaxy would appear similar to a
young galaxy that formed over a relatively short time-scale.
\citet{thomas-2005-621} claim that this degeneracy could be lifted by
relating observationally derived $\alpha$/Fe ratios with their
time-scales. Their simulations predict that an early-type galaxy of
mass $M_*/M_\odot \sim 10^{11.5}$, should have formed on a time-scale
of roughly 260~Myr. In their study, gaussians are used as a toy model
of star formation histories, whereas we use decaying
exponentials. Approximating their formation time-scale as the inverse
of $\gamma$ gives $\gamma \sim 3.8$~Gyr$^{-1}$. This suggests that
perhaps an even larger range of $\gamma$ is required to cover the full
range of physical star formation histories. It appears, though, that
the range we use works well for most of the galaxies. The distribution
of $\gamma$ values in the upper panel of Figure~\ref{fig:gamma} shows
a gradual decline towards the upper limit for the $0<\gamma<2$
fits. This is in contrast to a crowding of $\gamma$ values towards the
upper limit for the $0<\gamma<1$ fits. This suggests that, while some
galaxies may be better fit by models with $\gamma>2$, the range
$0<\gamma<2$ well describes the bulk of the LRG population.

We also check the consistency of our results by comparing the trends
shown by the 2D probability distributions of Figure~\ref{fig:twofinal}
with previous studies. Figure~12 of G05 shows similar distributions of
stellar metallicity and age for early-type galaxies in their
sample. We are interested in the upper mass bin, where
$\log(M/M_\odot) > 11$. They find that the high mass galaxies occupy a
region of parameter space centred on an age of about 8~Gyr and
metallicity of $[Z/H] = 0.1-0.2$. This is roughly consistent with the
age-metallicity plot in Figure~\ref{fig:twofinal}, which covers a
similar region. Both plots also show a similar age
metallicity-degeneracy trend, with a slope of $\Delta \log Z/\Delta
\log t = -1.05$~dex yr$^{-1}$ from our plot, compared with the value
of $-0.75$~dex yr$^{-1}$ found by G05. There are, however, some
differences. We find ages smaller than theirs by about 1--2~Gyr and
slightly higher metallicities. Our Figure~\ref{fig:comp} demonstrates
that these differences are due primarily to our different choice of
indices to fit and Monte-Carlo priors in the construction of the
library of template spectra.

Only a small percentage of the spectra are better fit by including a
burst on top of the continuos star formation. The probability that a
galaxy experienced a burst within the past 6~Gyr of its life is about
3 per cent, and only about 0.25 per cent within the past 2~Gyr. This
may in part be due to our selection criterion of avoiding spectra with
emission lines, but it is consistent with the small percentage of LRGs
showing evidence of recent star formation found by \citet{Rose06}. For
those galaxies better fit by the addition of a burst, the burst
contributes 20--40 per cent of the total stellar mass of the galaxy.

There have been a number of studies that show a distinct trend between
the properties of early-type galaxies and their velocity dispersions.
\citet{caldwell-2003-125} derive the ages and metallicities of 175
nearby, early-type galaxies using a combination of index-index
grids. They find a correlation between age and $\sigma$, with a
230~kms$^{-1}$ (the upper limit on their observations) galaxy expected
to have an age between 4 and 14~Gyr and a metallicity range of $-0.2 <
[Z/H] < 0.3$. Both of these results are broadly consistent with our
findings, and suggest that we are studying too narrow a range in
velocity dispersion or too homogeneous a population to detect any
significant trend.

The evolution in the median star formation rate we infer for the LRGs
agrees well with the semi-analytic galaxy formation predictions of
\citet{delucia-2006-366}. We agree less well on the formation times.
Comparing the times at which either 50 per cent or 80 per cent of
the stars in the LRGs were formed gives times shorter by about 2~Gyr
from the predictions of the semi-analytic models. We also find
metallicities about 0.15--0.2 dex larger than predicted.

In an attempt to understand the origin of the metallicity distribution
of the galaxies, we introduce a stochastic star-formation quenching
model for the metallicity evolution. In this model, metallicity
evolves either as a closed box or an accreting box, up until a time
when star formation abruptly ceases. The model is motivated by
quenching scenarios invoked to resolve the `anti-hierarchical'
formation problem, such as the shutting-off of gas cooling by AGN
activity. We find that while no model provides a perfect fit to the
data, the underlying measured distribution is well-described by both
the closed box and accreting box versions, provided a `cuspy' burst of
star-formation is adopted. The characteristic time-scale for quenching
the star formation is around $10^8$~yr, comparable to the estimated
lifetimes of luminous QSOs. The required effective yield values,
however, are quite different. For the closed box model, a yield of
about 0.26 is required. Such a high yield requires the enrichment to
be dominated by very massive stars. Such a possibility, however,
should perhaps not be discarded and is being considered in a newer set
of semi-analytic models (Lacey 2006, in preparation). For the
accreting box, a more moderate yield of about 0.10 is required. This
is still about twice the yield expected for the
\citet{2003PASP..115..763C} IMF for early stars, and again suggests an
IMF more strongly weighted towards higher masses. The dominant role
Type II supernovae would play in determining the abundances for such
IMFs is consistent with the $\alpha$-enhanced abundances inferred from
the absorption indices of LRGs.

The quenching model also predicts an anti-correlation between the ages
of the stars today and their metallicities, as early quenching will
result in older stars with smaller metallicities for a given formation
time. This provides a physical basis to the observed anti-correlation,
which may thus not soley be an artefact of the age-metallicity
degeneracy affecting population synthesis modelling. The relation,
however, will also depend on the range of formation times and decay
times for the star formation. Such effects are perhaps best accounted
for in semi-analytic galaxy formation models. Here we suggest that the
distribution of metallicity and the anti-correlation with age should
provide further constraints on such models.

\section*{Acknowledgments}

TB would like to thank the Royal Society of Edinburgh for their
support through a Robert Cormack Bequest Scholarship. TB would
also like to thank the School of Physics at the University of
Edinburgh for making this research possible.

\bibliographystyle{mn2e}
\bibliography{lrgs}

\begin{thebibliography}{}

\bibitem[\protect\citeauthoryear{{Adelman-McCarthy} \& {et
  al.}}{{Adelman-McCarthy} \& {et al.}}{2006}]{SDSSDR4-2006-162}
{Adelman-McCarthy} J.~K.,  {et al.} 2006, ApJS, 162, 38

\bibitem[\protect\citeauthoryear{{Althaus} \& {Benvenuto}}{{Althaus} \&
  {Benvenuto}}{1997}]{1997ApJ...477..313A}
{Althaus} L.~G.,  {Benvenuto} O.~G.,  1997, ApJ, 477, 313

\bibitem[\protect\citeauthoryear{{Balogh}, {Morris}, {Yee}, {Carlberg} \&
  {Ellingson}}{{Balogh} et~al.}{1999}]{balogh-1999-527}
{Balogh} M.~L.,  {Morris} S.~L.,  {Yee} H.~K.~C.,  {Carlberg} R.~G.,
  {Ellingson} E.,  1999, ApJ, 527, 54

\bibitem[\protect\citeauthoryear{{Baugh}, {Cole}, {Frenk} \& {Lacey}}{{Baugh}
  et~al.}{1998}]{1998ApJ...498..504B}
{Baugh} C.~M.,  {Cole} S.,  {Frenk} C.~S.,    {Lacey} C.~G.,  1998, ApJ, 498,
  504

\bibitem[\protect\citeauthoryear{{Binney} \& {Merrifield}}{{Binney} \&
  {Merrifield}}{1998}]{BM98}
{Binney} J.,  {Merrifield} M.,  1998, Galactic Astronomy.
Princeton University Press, Princeton, NJ

\bibitem[\protect\citeauthoryear{{Bloecker}}{{Bloecker}}{1995}]{1995A&A...299.%
.755B}
{Bloecker} T.,  1995, A\& A, 299, 755

\bibitem[\protect\citeauthoryear{{Blumenthal}, {Faber}, {Primack} \&
  {Rees}}{{Blumenthal} et~al.}{1984}]{1984Natur.311..517B}
{Blumenthal} G.~R.,  {Faber} S.~M.,  {Primack} J.~R.,    {Rees} M.~J.,  1984,
  Nature, 311, 517

\bibitem[\protect\citeauthoryear{{Boselli}, {Gavazzi}, {Donas} \&
  {Scodeggio}}{{Boselli} et~al.}{2001}]{2001AJ....121..753B}
{Boselli} A.,  {Gavazzi} G.,  {Donas} J.,    {Scodeggio} M.,  2001, AJ, 121,
  753

\bibitem[\protect\citeauthoryear{{Bower}, {Benson}, {Malbon}, {Helly}, {Frenk},
  {Baugh}, {Cole} \& {Lacey}}{{Bower} et~al.}{2006}]{2006MNRAS.370..645B}
{Bower} R.~G.,  {Benson} A.~J.,  {Malbon} R.,  {Helly} J.~C.,  {Frenk} C.~S.,
  {Baugh} C.~M.,  {Cole} S.,    {Lacey} C.~G.,  2006, MNRAS, 370, 645

\bibitem[\protect\citeauthoryear{{Bressan}, {Fagotto}, {Bertelli} \&
  {Chiosi}}{{Bressan} et~al.}{1993}]{bressan-1993-100}
{Bressan} A.,  {Fagotto} F.,  {Bertelli} G.,    {Chiosi} C.,  1993, A\&AS, 100,
  647

\bibitem[\protect\citeauthoryear{{Brinchmann}, {Charlot}, {White}, {Tremonti},
  {Kauffmann}, {Heckman} \& {Brinkmann}}{{Brinchmann}
  et~al.}{2004}]{2004MNRAS.351.1151B}
{Brinchmann} J.,  {Charlot} S.,  {White} S.~D.~M.,  {Tremonti} C.,  {Kauffmann}
  G.,  {Heckman} T.,    {Brinkmann} J.,  2004, MNRAS, 351, 1151

\bibitem[\protect\citeauthoryear{Bruzual \& Charlot}{Bruzual \&
  Charlot}{2003}]{bruzual-2003-344}
Bruzual G.,  Charlot S.,  2003, MNRAS, 344, 1000

\bibitem[\protect\citeauthoryear{{Bundy}, {Ellis} \& {Conselice}}{{Bundy}
  et~al.}{2005}]{2005ApJ...625..621B}
{Bundy} K.,  {Ellis} R.~S.,    {Conselice} C.~J.,  2005, ApJ, 625, 621

\bibitem[\protect\citeauthoryear{Bundy, Ellis, Conselice, Taylor, Cooper,
  Willmer, Weiner, Noeske \& Eisendardt}{Bundy et~al.}{2005}]{Bundy05}
Bundy K.,  Ellis R.~S.,  Conselice C.~J.,  Taylor J.~E.,  Cooper M.~C.,
  Willmer C. N.~A.,  Weiner B.~J.,  Noeske K.~G.,    Eisendardt P. R.~M., ,
  2005, preprint, {astro-ph}/0512465

\bibitem[\protect\citeauthoryear{{Burstein}, {Faber}, {Gaskell} \&
  {Krumm}}{{Burstein} et~al.}{1984}]{burstein-1984-287}
{Burstein} D.,  {Faber} S.~M.,  {Gaskell} C.~M.,    {Krumm} N.,  1984, ApJ,
  287, 586

\bibitem[\protect\citeauthoryear{{Caldwell}, {Rose} \& {Concannon}}{{Caldwell}
  et~al.}{2003}]{caldwell-2003-125}
{Caldwell} N.,  {Rose} J.~A.,    {Concannon} K.~D.,  2003, AJ, 125, 2891

\bibitem[\protect\citeauthoryear{{Calzetti}, {Kinney} \&
  {Storchi-Bergmann}}{{Calzetti} et~al.}{1994}]{calzetti-1994-429}
{Calzetti} D.,  {Kinney} A.~L.,    {Storchi-Bergmann} T.,  1994, ApJ, 429, 582

\bibitem[\protect\citeauthoryear{{Cardelli}, {Clayton} \& {Mathis}}{{Cardelli}
  et~al.}{1989}]{cardelli-1989-345}
{Cardelli} J.~A.,  {Clayton} G.~C.,    {Mathis} J.~S.,  1989, ApJ, 345, 245

\bibitem[\protect\citeauthoryear{{Cardiel}, {Gorgas}, {Cenarro} \&
  {Gonzalez}}{{Cardiel} et~al.}{1998}]{cardiel-1998-127}
{Cardiel} N.,  {Gorgas} J.,  {Cenarro} J.,    {Gonzalez} J.~J.,  1998, A\& AS,
  127, 597

\bibitem[\protect\citeauthoryear{{Chabrier}}{{Chabrier}}{2003}]{2003PASP..115.%
.763C}
{Chabrier} G.,  2003, PASP, 115, 763

\bibitem[\protect\citeauthoryear{{Chabrier} \& {Baraffe}}{{Chabrier} \&
  {Baraffe}}{1997}]{1997A&A...327.1039C}
{Chabrier} G.,  {Baraffe} I.,  1997, A\& A, 327, 1039

\bibitem[\protect\citeauthoryear{Cimatti, Daddi \& Renzini}{Cimatti
  et~al.}{2006}]{Cimatti06}
Cimatti A.,  Daddi E.,    Renzini A., , 2006, preprint, {astro-ph}/0605353

\bibitem[\protect\citeauthoryear{{Cole}, {Aragon-Salamanca}, {Frenk}, {Navarro}
  \& {Zepf}}{{Cole} et~al.}{1994}]{1994MNRAS.271..781C}
{Cole} S.,  {Aragon-Salamanca} A.,  {Frenk} C.~S.,  {Navarro} J.~F.,    {Zepf}
  S.~E.,  1994, MNRAS, 271, 781

\bibitem[\protect\citeauthoryear{{Conselice}}{{Conselice}}{2006}]{2006ApJ...63%
8..686C}
{Conselice} C.~J.,  2006, ApJ, 638, 686

\bibitem[\protect\citeauthoryear{{Cowie}, {Songaila}, {Hu} \& {Cohen}}{{Cowie}
  et~al.}{1996}]{1996AJ....112..839C}
{Cowie} L.~L.,  {Songaila} A.,  {Hu} E.~M.,    {Cohen} J.~G.,  1996, AJ, 112,
  839

\bibitem[\protect\citeauthoryear{{Davies}, {Sadler} \& {Peletier}}{{Davies}
  et~al.}{1993}]{1993MNRAS.262..650D}
{Davies} R.~L.,  {Sadler} E.~M.,    {Peletier} R.~F.,  1993, MNRAS, 262, 650

\bibitem[\protect\citeauthoryear{{Davis}, {Efstathiou}, {Frenk} \&
  {White}}{{Davis} et~al.}{1985}]{1985ApJ...292..371D}
{Davis} M.,  {Efstathiou} G.,  {Frenk} C.~S.,    {White} S.~D.~M.,  1985, ApJ,
  292, 371

\bibitem[\protect\citeauthoryear{{De Lucia}, {Springel}, {White}, {Croton} \&
  {Kauffmann}}{{De Lucia} et~al.}{2006}]{delucia-2006-366}
{De Lucia} G.,  {Springel} V.,  {White} S.~D.~M.,  {Croton} D.,    {Kauffmann}
  G.,  2006, MNRAS, 366, 499

\bibitem[\protect\citeauthoryear{{Dekel} \& {Birnboim}}{{Dekel} \&
  {Birnboim}}{2006}]{2006MNRAS.368....2D}
{Dekel} A.,  {Birnboim} Y.,  2006, MNRAS, 368, 2

\bibitem[\protect\citeauthoryear{{Drory}, {Salvato}, {Gabasch}, {Bender},
  {Hopp}, {Feulner} \& {Pannella}}{{Drory} et~al.}{2005}]{2005ApJ...619L.131D}
{Drory} N.,  {Salvato} M.,  {Gabasch} A.,  {Bender} R.,  {Hopp} U.,  {Feulner}
  G.,    {Pannella} M.,  2005, ApJ, 619, L131

\bibitem[\protect\citeauthoryear{{Eisenstein} \& et al.}{{Eisenstein} \&
  et~al.}{2001}]{eisenstein-2001-122}
{Eisenstein} D.~J.,  et al. 2001, AJ, 122, 2267

\bibitem[\protect\citeauthoryear{{Eisenstein} \& et al.}{{Eisenstein} \&
  et~al.}{2003}]{2003ApJ...585..694E}
{Eisenstein} D.~J.,  et al. 2003, ApJ, 585, 694

\bibitem[\protect\citeauthoryear{{Faber}}{{Faber}}{1973}]{1973A&AS...10..201F}
{Faber} S.~M.,  1973, A\&AS, 10, 201

\bibitem[\protect\citeauthoryear{{Fioc} \& {Rocca-Volmerange}}{{Fioc} \&
  {Rocca-Volmerange}}{1997}]{PEGASE-1997-326}
{Fioc} M.,  {Rocca-Volmerange} B.,  1997, A\&A, 326, 950

\bibitem[\protect\citeauthoryear{{Fontana}, {Pozzetti}, {Donnarumma},
  {Renzini}, {Cimatti}, {Zamorani}, {Menci}, {Daddi}, {Giallongo}, {Mignoli},
  {Perna}, {Salimbeni}, {Saracco}, {Broadhurst}, {Cristiani}, {D'Odorico} \&
  {Gilmozzi}}{{Fontana} et~al.}{2004}]{2004A&A...424...23F}
{Fontana} A.,  {Pozzetti} L.,  {Donnarumma} I.,  {Renzini} A.,  {Cimatti} A.,
  {Zamorani} G.,  {Menci} N.,  {Daddi} E.,  {Giallongo} E.,  {Mignoli} M.,
  {Perna} C.,  {Salimbeni} S.,  {Saracco} P.,  {Broadhurst} T.,  {Cristiani}
  S.,  {D'Odorico} S.,    {Gilmozzi} R.,  2004, A\&A, 424, 23

\bibitem[\protect\citeauthoryear{{Gallazzi}, {Charlot}, {Brinchmann}, {White}
  \& {Tremonti}}{{Gallazzi} et~al.}{2005}]{gallazzi-2005-362}
{Gallazzi} A.,  {Charlot} S.,  {Brinchmann} J.,  {White} S.~D.~M.,
  {Tremonti} C.~A.,  2005, MNRAS, 362, 41

\bibitem[\protect\citeauthoryear{{Gavazzi}, {Boselli}, {Pedotti}, {Gallazzi} \&
  {Carrasco}}{{Gavazzi} et~al.}{2002}]{2002A&A...396..449G}
{Gavazzi} G.,  {Boselli} A.,  {Pedotti} P.,  {Gallazzi} A.,    {Carrasco} L.,
  2002, A\&A, 396, 449

\bibitem[\protect\citeauthoryear{{Groenewegen} \& {de Jong}}{{Groenewegen} \&
  {de Jong}}{1993}]{1993A&A...267..410G}
{Groenewegen} M.~A.~T.,  {de Jong} T.,  1993, A\& A, 267, 410

\bibitem[\protect\citeauthoryear{{Heyl}, {Colless}, {Ellis} \&
  {Broadhurst}}{{Heyl} et~al.}{1997}]{1997MNRAS.285..613H}
{Heyl} J.,  {Colless} M.,  {Ellis} R.~S.,    {Broadhurst} T.,  1997, MNRAS,
  285, 613

\bibitem[\protect\citeauthoryear{{Hirschi}, {Meynet} \& {Maeder}}{{Hirschi}
  et~al.}{2005}]{2005A&A...433.1013H}
{Hirschi} R.,  {Meynet} G.,    {Maeder} A.,  2005, A\& A, 433, 1013

\bibitem[\protect\citeauthoryear{{Jones} \& {Worthey}}{{Jones} \&
  {Worthey}}{1995}]{jones-1995-446}
{Jones} L.~A.,  {Worthey} G.,  1995, ApJ, 446, L31

\bibitem[\protect\citeauthoryear{{Kauffmann}}{{Kauffmann}}{1996}]{1996MNRAS.28%
1..487K}
{Kauffmann} G.,  1996, MNRAS, 281, 487

\bibitem[\protect\citeauthoryear{{Kauffmann} \& {Charlot}}{{Kauffmann} \&
  {Charlot}}{1998}]{1998MNRAS.294..705K}
{Kauffmann} G.,  {Charlot} S.,  1998, MNRAS, 294, 705

\bibitem[\protect\citeauthoryear{Kauffmann, Heckman, White, Charlot, Tremonti,
  Brinchmann, Bruzual, Peng, Seibert, Bernardi, Blanton, Brinkmann, Castander,
  Csabai, Fukugita, Ivezic, Munn, Nichol, Padmanabhan, Thakar, Weinberg \&
  York}{Kauffmann et~al.}{2003}]{kauffmann-2003-341}
Kauffmann G.,  Heckman T.~M.,  White S. D.~M.,  Charlot S.,  Tremonti C.,
  Brinchmann J.,  Bruzual G.,  Peng E.~W.,  Seibert M.,  Bernardi M.,  Blanton
  M.,  Brinkmann J.,  Castander F.,  Csabai I.,  Fukugita M.,  Ivezic Z.,  Munn
  J.,  Nichol R.,  Padmanabhan N.,  Thakar A.,  Weinberg D.,    York D.,  2003,
  MNRAS, 341, 33

\bibitem[\protect\citeauthoryear{{Kelson}, {Illingworth}, {Franx} \& {van
  Dokkum}}{{Kelson} et~al.}{2006}]{kelson-2006-642}
{Kelson} D.~D.,  {Illingworth} G.~D.,  {Franx} M.,    {van Dokkum} P.~G., ,
  2006, preprint, {astro-ph}/0606642

\bibitem[\protect\citeauthoryear{{Kormendy} \& {Djorgovski}}{{Kormendy} \&
  {Djorgovski}}{1989}]{1989ARA&A..27..235K}
{Kormendy} J.,  {Djorgovski} S.,  1989, ARA\& A, 27, 235

\bibitem[\protect\citeauthoryear{{Korn}, {Maraston} \& {Thomas}}{{Korn}
  et~al.}{2005}]{korn-2005-438}
{Korn} A.~J.,  {Maraston} C.,    {Thomas} D.,  2005, A\&A, 438, 685

\bibitem[\protect\citeauthoryear{{Kroupa}}{{Kroupa}}{2001}]{2001MNRAS.322..231%
K}
{Kroupa} P.,  2001, MNRAS, 322, 231

\bibitem[\protect\citeauthoryear{{Le Borgne}, {Bruzual}, {Pell{\'o}}, {Lan{\c
  c}on}, {Rocca-Volmerange}, {Sanahuja}, {Schaerer}, {Soubiran} \&
  {V{\'{\i}}lchez-G{\'o}mez}}{{Le Borgne} et~al.}{2003}]{2003A&A...402..433L}
{Le Borgne} J.-F.,  {Bruzual} G.,  {Pell{\'o}} R.,  {Lan{\c c}on} A.,
  {Rocca-Volmerange} B.,  {Sanahuja} B.,  {Schaerer} D.,  {Soubiran} C.,
  {V{\'{\i}}lchez-G{\'o}mez} R.,  2003, A\& A, 402, 433

\bibitem[\protect\citeauthoryear{Lucia \& Blaizot}{Lucia \&
  Blaizot}{2006}]{DeLucia06}
Lucia G.~D.,  Blaizot J., , 2006, preprint, {astro-ph}/0606519

\bibitem[\protect\citeauthoryear{{Maeder}}{{Maeder}}{1992}]{1992A&A...264..105%
M}
{Maeder} A.,  1992, A\& A, 264, 105

\bibitem[\protect\citeauthoryear{Murphy}{Murphy}{2003}]{murphy-2003-thesis}
Murphy T.,  2003, PhD thesis, University of Edinburgh

\bibitem[\protect\citeauthoryear{Murphy \& Meiksin}{Murphy \&
  Meiksin}{2004}]{murphy-2004-351}
Murphy T.,  Meiksin A.,  2004, MNRAS, 351, 1430

\bibitem[\protect\citeauthoryear{{Peebles}}{{Peebles}}{1983}]{1983ApJ...274...%
.1P}
{Peebles} P.~J.~E.,  1983, ApJ, 274, 1

\bibitem[\protect\citeauthoryear{{Peebles} \& {Dicke}}{{Peebles} \&
  {Dicke}}{1968}]{1968ApJ...154..891P}
{Peebles} P.~J.~E.,  {Dicke} R.~H.,  1968, ApJ, 154, 891

\bibitem[\protect\citeauthoryear{{Pickles}}{{Pickles}}{1985}]{1985ApJ...296..3%
40P}
{Pickles} A.~J.,  1985, ApJ, 296, 340

\bibitem[\protect\citeauthoryear{{Pozzetti}, {Cimatti}, {Zamorani}, {Daddi},
  {Menci}, {Fontana}, {Renzini}, {Mignoli}, {Poli}, {Saracco}, {Broadhurst},
  {Cristiani}, {D'Odorico}, {Giallongo} \& {Gilmozzi}}{{Pozzetti}
  et~al.}{2003}]{2003A&A...402..837P}
{Pozzetti} L.,  {Cimatti} A.,  {Zamorani} G.,  {Daddi} E.,  {Menci} N.,
  {Fontana} A.,  {Renzini} A.,  {Mignoli} M.,  {Poli} F.,  {Saracco} P.,
  {Broadhurst} T.,  {Cristiani} S.,  {D'Odorico} S.,  {Giallongo} E.,
  {Gilmozzi} R.,  2003, A\&A, 402, 837

\bibitem[\protect\citeauthoryear{{Roseboom} \& {et al.}}{{Roseboom} \& {et
  al.}}{2006}]{Rose06}
{Roseboom} I.~G.,  {et al.}, 2006, preprint, {astro-ph}/0609178

\bibitem[\protect\citeauthoryear{{Sawicki}, {Lin} \& {Yee}}{{Sawicki}
  et~al.}{1997}]{1997AJ....113....1S}
{Sawicki} M.~J.,  {Lin} H.,    {Yee} H.~K.~C.,  1997, AJ, 113, 1

\bibitem[\protect\citeauthoryear{{Schlegel}, {Finkbeiner} \&
  {Davis}}{{Schlegel} et~al.}{1998}]{schlegel-1998-500}
{Schlegel} D.~J.,  {Finkbeiner} D.~P.,    {Davis} M.,  1998, ApJ, 500, 525

\bibitem[\protect\citeauthoryear{{Schoenberner}}{{Schoenberner}}{1983}]{1983Ap%
J...272..708S}
{Schoenberner} D.,  1983, ApJ, 272, 708

\bibitem[\protect\citeauthoryear{{Stoughton} \& et al.}{{Stoughton} \&
  et~al.}{2002}]{SDSSEDR-2002-123}
{Stoughton} C.,  et al. 2002, AJ, 123, 485

\bibitem[\protect\citeauthoryear{{Talbot} Jr. \& {Arnett}}{{Talbot} \&
  {Arnett}}{1971}]{talbot-1971-170}
{Talbot} Jr. R.~J.,  {Arnett} W.~D.,  1971, ApJ, 170, 409

\bibitem[\protect\citeauthoryear{{Thomas}, {Maraston} \& {Bender}}{{Thomas}
  et~al.}{2003}]{thomas-2003-339}
{Thomas} D.,  {Maraston} C.,    {Bender} R.,  2003, MNRAS, 339, 897

\bibitem[\protect\citeauthoryear{{Thomas}, {Maraston}, {Bender} \& {Mendes de
  Oliveira}}{{Thomas} et~al.}{2005}]{thomas-2005-621}
{Thomas} D.,  {Maraston} C.,  {Bender} R.,    {Mendes de Oliveira} C.,  2005,
  ApJ, 621, 673

\bibitem[\protect\citeauthoryear{{Trager}, {Faber}, {Worthey} \&
  {Gonz{\'a}lez}}{{Trager} et~al.}{2000a}]{2000AJ....120..165T}
{Trager} S.~C.,  {Faber} S.~M.,  {Worthey} G.,    {Gonz{\'a}lez} J.~J.,  2000a,
  AJ, 120, 165

\bibitem[\protect\citeauthoryear{{Trager}, {Faber}, {Worthey} \&
  {Gonz{\'a}lez}}{{Trager} et~al.}{2000b}]{trager-2000-119}
{Trager} S.~C.,  {Faber} S.~M.,  {Worthey} G.,    {Gonz{\'a}lez} J.~J.,  2000b,
  AJ, 119, 1645

\bibitem[\protect\citeauthoryear{{Treu}, {Stiavelli}, {Casertano}, {M{\o}ller}
  \& {Bertin}}{{Treu} et~al.}{2002}]{2002ApJ...564L..13T}
{Treu} T.,  {Stiavelli} M.,  {Casertano} S.,  {M{\o}ller} P.,    {Bertin} G.,
  2002, ApJ, 564, L13

\bibitem[\protect\citeauthoryear{{Visvanathan} \& {Sandage}}{{Visvanathan} \&
  {Sandage}}{1977}]{1977ApJ...216..214V}
{Visvanathan} N.,  {Sandage} A.,  1977, ApJ, 216, 214

\bibitem[\protect\citeauthoryear{{White} \& {Rees}}{{White} \&
  {Rees}}{1978}]{1978MNRAS.183..341W}
{White} S.~D.~M.,  {Rees} M.~J.,  1978, MNRAS, 183, 341

\bibitem[\protect\citeauthoryear{{Worthey}, {Faber} \& {Gonzalez}}{{Worthey}
  et~al.}{1992}]{1992ApJ...398...69W}
{Worthey} G.,  {Faber} S.~M.,    {Gonzalez} J.~J.,  1992, ApJ, 398, 69

\bibitem[\protect\citeauthoryear{{Worthey}, {Faber}, {Gonzalez} \&
  {Burstein}}{{Worthey} et~al.}{1994}]{worthey-1994-94}
{Worthey} G.,  {Faber} S.~M.,  {Gonzalez} J.~J.,    {Burstein} D.,  1994, ApJS,
  94, 687

\bibitem[\protect\citeauthoryear{{Worthey} \& {Ottaviani}}{{Worthey} \&
  {Ottaviani}}{1997}]{worthey-1997-111}
{Worthey} G.,  {Ottaviani} D.~L.,  1997, ApJS, 111, 377

\bibitem[\protect\citeauthoryear{{Yamada}, {Arimoto}, {Vazdekis} \&
  {Peletier}}{{Yamada} et~al.}{2006}]{2006ApJ...637..200Y}
{Yamada} Y.,  {Arimoto} N.,  {Vazdekis} A.,    {Peletier} R.~F.,  2006, ApJ,
  637, 200

\bibitem[\protect\citeauthoryear{{York} \& {et al.}}{{York} \& {et
  al.}}{2000}]{2000AJ....120.1579Y}
{York} D.~G.,  {et al.} 2000, AJ, 120, 1579

\bibitem[\protect\citeauthoryear{{Yu} \& {Tremaine}}{{Yu} \&
  {Tremaine}}{2002}]{2002MNRAS.335..965Y}
{Yu} Q.,  {Tremaine} S.,  2002, MNRAS, 335, 965

\end{thebibliography}

\label{lastpage}

\end{document}